\documentclass[a4paper,11pt]{article}
\pdfoutput=1

\usepackage{jheppub}
\usepackage{multirow, graphicx,amssymb,url,mathrsfs,amsmath}
\usepackage{wrapfig,boxedminipage,setspace,subfigure,epsfig}
\usepackage{amsxtra,amstext,latexsym,dsfont,amsfonts}

\newcommand{\td}{\text{d}}
\newcommand{\el}{\ell_{\text{AdS}}}

\newcommand{\kyr}[1]{{\color{black}{#1}}}
\newcommand{\kyb}[1]{{\color{black}{#1}}}

\begin{document}
\title{\boldmath Surface Counterterms and Regularized Holographic Complexity}
\author[a]{Run-Qiu Yang,}
\author[b]{Chao Niu,}
\author[b]{Keun-Young Kim}

\emailAdd{aqiu@kias.re.kr}
\emailAdd{chaoniu09@gmail.com}
\emailAdd{fortoe@gist.ac.kr}

\affiliation[a]{Quantum Universe Center, Korea Institute for Advanced Study, Seoul 130-722, Korea}
\affiliation[b]{ School of Physics and Chemistry, Gwangju Institute of Science and Technology,
Gwangju 61005, Korea
}

\abstract{The holographic complexity is UV divergent.
As a finite complexity, we propose a ``regularized complexity''  by employing a similar method to the holographic renormalization.
We add codimension-two boundary counterterms which do not contain any boundary stress tensor information. It means that we subtract only non-dynamic background and all the dynamic information of holographic complexity is contained in the regularized part. \kyr{After showing the general counterterms for both CA and CV conjectures in holographic spacetime dimension 5 and less, we give concrete examples: the BTZ black holes and the four and five dimensional Schwarzschild AdS black holes. We propose how to obtain the counterterms in higher spacetime dimensions \kyb{and show explicit formulas only for some special cases with enough symmetries.}} We also compute the complexity of formation by using the regularized complexity.}

\maketitle
\flushbottom

\noindent

\section{Introduction}\label{intro}

\begin{figure}
  \centering
  \includegraphics[width=.4\textwidth]{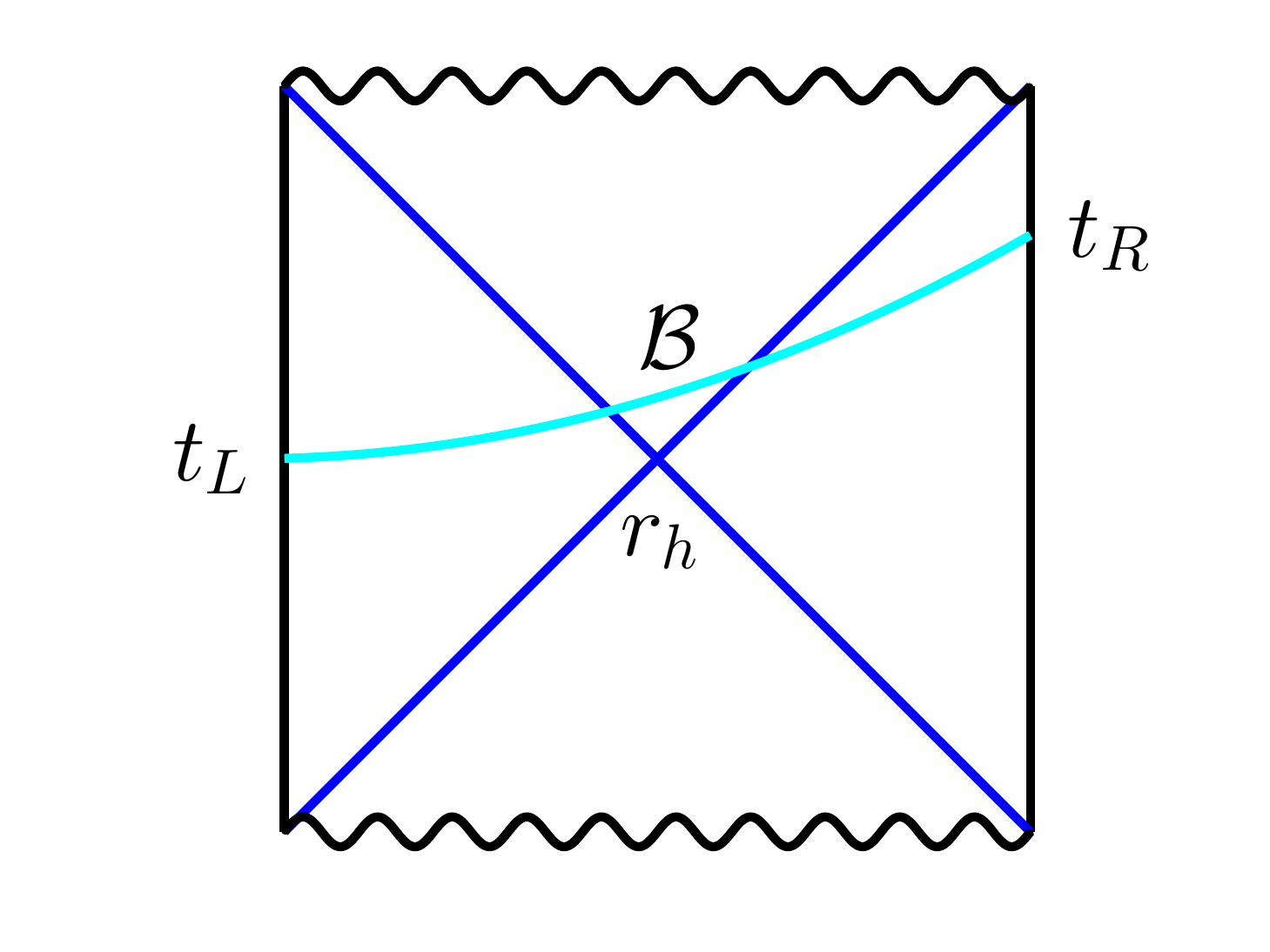}
   \includegraphics[width=.4\textwidth]{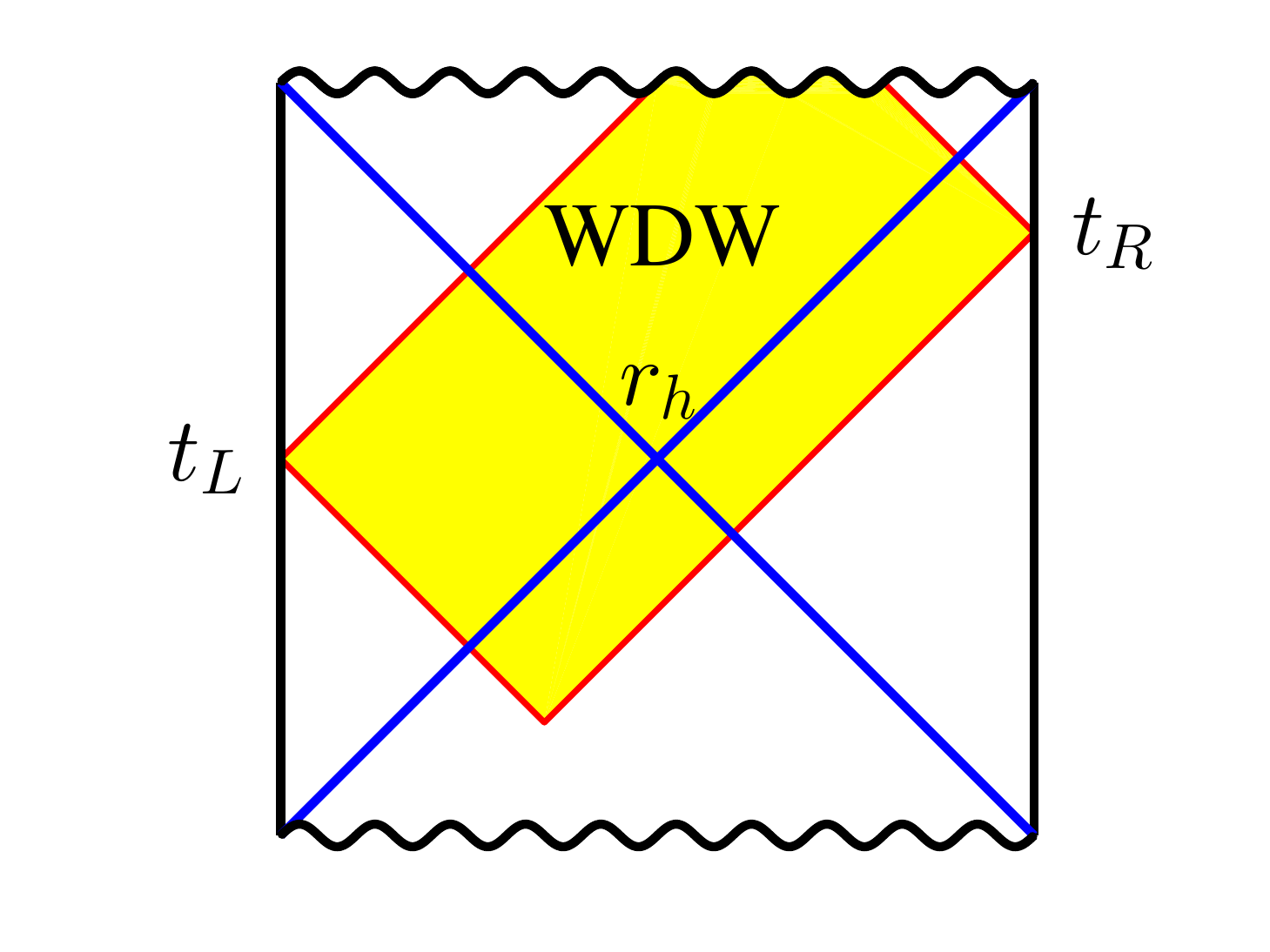}
  \caption{Penrose diagram for Schwarzschild AdS black hole and complexity in two conjectures.
 At the two boundaries of the black hole, $t_L$ and $t_R$ stand for two states dual to the  states in TFD. $r_h$ is the horizon radius. At the left panel, $\mathcal{B}$ is the maximum codimension-one surface connecting $t_L$ and $t_R$. At the right panel, the yellow region with its boundary is the WDW patch, which is the closure (inner region with the boundary) of all space-like codimension-one surfaces connecting $t_L$ and $t_R$.  }\label{TFDAdS}
\end{figure}

In recent years, the quantum entanglement and information gave us new viewpoints to  see quantum gravity and black holes. One of the very interesting results in this aspect is the  quantum complexity and its gravity dual description. Roughly speaking, the quantum complexity characterizes how difficult it is to obtain a particular quantum state from an appointed reference state.  In a discrete system, such as a quantum logic circuit, it's the minimal number of simple gates from the reference state to a particular state \cite{2008arXiv0804.3401W,2014arXiv1401.3916G,2012RPPh...75b2001O}. The quantum entanglement has also been found to play an important role in the quantum gravity, especially for the study on the AdS/CFT correspondence. While most of recent works have paid attention to the holographic entanglement entropy \cite{Ryu:2006bv,Ryu:2006ef}, quantum complexity in gravity was studied in \cite{Susskind:2014rva,Stanford:2014jda,Susskind:2014moa,Brown:2015bva,Brown:2015lvg}: by paying attention to the growth of the Einstein-Rosen bridge the authors found a connection between AdS black hole and quantum complexity in the dual boundary conformal field theory (CFT). In this study, they consider the eternal AdS black holes, which are dual to thermofield double (TFD) state~\cite{Maldacena:2001kr}
%
\begin{equation}\label{TFD1}
  |\text{TFD}\rangle:=Z^{-1/2}\sum_\alpha\exp[-E_\alpha/(2T)]|E_\alpha\rangle_L |E_\alpha\rangle_R \,.
\end{equation}
The states $|E_\alpha\rangle_L$ and $|E_\alpha\rangle_R$ are defined in the two copy CFTs at the two boundaries of the eternal AdS black hole (see Fig. \ref{TFDAdS}) and $T$ is the temperature. {With the Hamiltonians $H_L$ and $H_R$ at the left and right dual CFTs, the time evolution of a TFD state
\begin{equation}\label{timesate1}
  |\psi(t_L,t_R)\rangle:=e^{-i(t_LH_L+t_RH_R)}|\text{TFD}\rangle\,
\end{equation}
can be characterized by the codimension-two surface at  fixed times $t=t_L$ and $t=t_R$ at the two boundaries of the AdS black hole~\cite{Maldacena:2001kr,Brown:2015lvg}.} There are two proposals to compute the complexity of $|\psi(t_L,t_R)\rangle$ state holographically: CV(complexity=volume) conjecture and CA(complexity= action) conjecture.

\vspace{0.3cm}
The CV conjecture~\cite{Stanford:2014jda,Alishahiha:2015rta} states that the complexity of {$|\psi(t_L,t_R)\rangle$} at the boundary CFT is proportional to the maximal volume of the space-like codimension-one surface which connects the codimension-two surfaces denoted by $t_L$ and $t_R$, i.e.
\begin{equation}\label{CV}
  \mathcal{C}_V=\max_{\partial \Sigma=t_L\cup t_R}\left[\frac{V(\Sigma)}{G_N \ell}\right] \,,
\end{equation}
where $G_N$ is the Newton's constant. $\Sigma$ is all the possible space-like codimension-one surfaces which connect $t_L$ and $t_R$ and $\ell$ is a length scale associated with the bulk geometry such as horizon radius or AdS radius and so on.
This conjecture satisfies some  properties of the quantum complexity. However, there is an ambiguity coming from the choice of a length scale $\ell$.

\vspace{0.3cm}
This unsatisfactory feature motivated the second conjecture: CA conjecture \cite{Brown:2015bva,Brown:2015lvg}. In this conjecture, the complexity of  a {$|\psi(t_L,t_R)\rangle$} is dual to the action in the Wheeler-DeWitt (WDW) patch associated with $t_L$ and $t_R$, i.e.
\begin{equation}\label{CA}
  \mathcal{C}_A=\frac{I_{\text{WDW}}}{\pi\hbar}.
\end{equation}
The WDW patch associated with $t_L$ and $t_R$ is the collection of all space-like surface connecting $t_L$ and $t_R$ with the null sheets coming from $t_L$ and $t_R$. More precisely it is the domain of dependence of any space-like surface connecting $t_L$ and $t_R$ (see the right panel of Fig. \ref{TFDAdS} as an example). This conjecture has some advantages compared with the CV conjecture. For example, it has no free parameter and can satisfy Lloyd's complexity growth bound in very general cases \cite{Lloyd2000,Cai:2016xho,Yang:2016awy}. However, the CA conjecture has its own obstacle in computing the action: it involves null boundaries and joint terms. Recently, this problem has been overcome by carefully analyzing the boundary term in null boundary \cite{Parattu:2015gga,Lehner:2016vdi}.

As both the CV and CA conjectures involve the integration over infinite region, the complexity computed by  the Eqs. \eqref{CV} and \eqref{CA} are divergent. The divergences appearing in the CV and CA conjectures are similar to the one in the holographic entanglement entropy. It was shown that the coefficients of all the divergent terms can be written as the local integration of boundary geometry \cite{Carmi:2016wjl,Reynolds:2016rvl}, which is independent of the bulk stress tensor. This result gives a clear physical meaning of the divergences in the holographic complexity: they come from the UV vacuum structure at a given time slice and stand for the vacuum CFT's contribution to the complexity. One interesting thing is to consider the contribution of excited state or thermal state to the complexity.  As the divergent parts of the holographic complexity is fixed by the boundary geometry, the contribution of matter fields and temperature can only appear in the finite term of the complexity. This gives us a strong motivation to study how to obtain the finite term in the complexity.

The first work regarding this finite quantity is the ``complexity of formation'' \cite{Chapman:2016hwi}, which is defined by the difference of the complexity in a particular black hole space time and a reference vacuum AdS space-time. By choosing a suitable vacuum space-time, we can obtain a finite complexity of formation. However, there are two somewhat ambiguous aspects in using ``complexity of formation'' to study the finite term of complexity. First, we need to appoint additional space-time as the reference vacuum background. In general cases, it will not be obvious  how to choose the reference vacuum space-time. For example, in Ref.  \cite{Chapman:2016hwi}, the reference vacuum space-time for the BTZ black hole is not the naive limit of setting mass $M=0$. Second, to make the computation about the difference of complexity at the finite cut-off between two space-times meaningful, we need to appoint a special coordinate and apply this coordinate to both space-times. For example, in the Ref.  \cite{Chapman:2016hwi}, the holographic complexity of two space-time at the finite cut-off is computed in Fefferman-Graham coordinate \cite{Graham:1999pm,2007arXiv0710.0919F}.  It will be better if we can compute the complexity without referring to a specific coordinate system.

As the Refs. \cite{Carmi:2016wjl,Reynolds:2016rvl} have shown that the divergent terms have some universal structures, a naive consideration is that, we can separate the divergent term and just discard them.
However, this may give a coordinate dependent result as we shows in the section \ref{exmp1}.
In this paper, we will propose another method to obtain the finite term of the complexity, which we will call ``regularized complexity''.   \kyr{Colsely following the method of the holographic renormalization \cite{Emparan:1999pm,deHaro:2000vlm,Skenderis:2002wp,Bianchi:2001kw} we will add  codimension-two surface counterterms for a given dimension $d+1$,\footnote{\kyr{For holographic renormalization of entanglement entropy, we refer to \cite{Hertzberg:2010uv,Liu:2012eea,Taylor:2016aoi}. In particular, our method is similar to  \cite{Taylor:2016aoi}. } }
\begin{equation}\label{rr2}
  V_{\text{ct}}=\el \int_B\td^{d-1}x \ \sqrt{\sigma}\,  \sum_{n=0}^{[\frac{d-1}2]}  \el^{2n}\ F_{V}^{(2n)}(d, R_{\mu\nu}, g_{\mu\nu}, \sigma_{ij},K_{ij} )\,, 
\end{equation}
\begin{equation}\label{rr1}
   I_{\text{ct}}
   =  \frac{1}{G_N} \int_B\td^{d-1}x \ \sqrt{\sigma}\,  \sum_{n=0}^{[\frac{d-1}2]}  \el^{2n}\ F_{A}^{(2n)}(d, R_{\mu\nu}, g_{\mu\nu}, \sigma_{ij},K_{ij} ) \,, 
\end{equation}
to the complexity formula in the CV and CA conjectures \eqref{CV} and \eqref{CA}  respectively.\footnote{In this paper, the  capital latin letters $I,J,\cdots$ run from 0 to $d$, which stand for the all coordinates and $x^d=z$. The Greek indices $\mu,\nu,\cdots$ run from 0 to $d-1$, which stand for the local coordinate at the fixed $z$ surface and $x^0=t$. The little latin letters $i,j,\cdots$ run from 1 to $d-1$, which stand for the local coordinates at the fixed $z$ and $t$ surface.}
Here $B$ is the codimension-two surface of given time $t=t_L$ or $t_R$ at the cut-off boundaries. $\el$ is the radius of the AdS space. $g_{\mu\nu}$ is the induced metric at the cut-off boundaries, $R_{\mu\nu}$ is the Ricci tensor from $g_{\mu\nu}$, $\sigma_{ij}$ is the induced metric of the time slice $t_L$ or $t_R$ and $K_{ij}$ is the extrinsic curvature of the time slice $t_L$ or $t_R$ embedded into the boundaries.  $F_{V}^{(2n)}$ and $F_A^{(2n)}$ are invariant combinations of $R_{\mu\nu}, g_{\mu\nu}, \sigma_{ij}$ and $K_{ij}$ with a mass dimension $2n$, so $V_{\text{ct}}$ is of volume dimension $d$ and $I_{\text{ct}}$ is dimensionless. The concrete form of $F_{V}^{(2n)}$ and $F_A^{(2n)}$ will be determined based on the divergent structure developed in \cite{Carmi:2016wjl,Reynolds:2016rvl}.  When the bulk dimension is even ($d$ is odd) a logarithmic divergence appears, and $F_{V}^{(d-1)}$ and $F_A^{(d-1)}$  should be understood as a counterterm for the logarithmic divergence. The counter terms are determined by the boundary metric alone and do not contain any boundary stress tensor information.

}

The procedure to obtain the regularized complexity is  similar to holographic renormalization. However, there are two  differences. First, the surface counterterms we will show are the codimension-two surface at the boundary rather than the codimension-one surface.  Because the complexity, as shown in the Fig. \ref{TFDAdS}, is defined by the time slices denoted by $t_L$ and $t_R$, which are codimension-two surfaces, it is natural that the surface counterterms should be expressed as the geometric quantities of these codimension-two surfaces. Second,  the surface counterterms can contain the extrinsic geometrical quantities of the codimension-two boundary rather than only the intrinsic geometrical quantities unlike in renormalizing free energy. One  reason for this difference is that free energy involves the equations of motion and we need to keep the equations of motion invariant when we renormalize the free energy but complexity has no directly relationship with the equation of motion.

The organization of this paper is as follows. In section \ref{surf1}, we will give the surface counterterms for both CA and CV conjectures. We first show an example how the coordinate dependence appears if we just discard the divergent terms, which will give the inspiration on how to construct the surface counterterms. Then we will explicitly give the  minimal subtraction counterterms both for CA and CV conjectures up to the bulk dimension $d+1\leq5$. In the sections \ref{exmplBTZ} and \ref{exmplSAdS}, we will use our surface counterterms to compute the regularized complexity for the  BTZ black holes and Schwarzschild AdS  black holes  for both CA and CV conjectures. A summary will be found in section~\ref{Summary}.

\section{Surface counterterms and regularized complexity} \label{surf1}
\subsection{Coordinate dependence in discarding divergent terms}\label{exmp1}

To regularize the complexity  we may try the same method as the entanglement entropy case, for example, in Refs.~\cite{Albash:2012pd,Albash:2010mv,Ling:2015dma} i.e. find out the divergent behavior and then just discard all the divergent terms. However, in the following example for the CV conjecture, we will show such a method is a coordinate-dependent so ambiguous.   Such coordinate dependence can appear also in the CA conjecture, subregion complexity,  and in the entanglement entropy, if we just naively discard the divergent terms.

Let us first consider a Schwarzschild AdS$_4$ black brane geometry
\begin{equation}\label{SAdSmetric1}
  \td s^2=-r^2f(r)\td t^2+\frac{\td r^2}{r^2f(r)}+r^2(\td x^2+\td y^2) \,,
\end{equation}
with
\begin{equation}\label{SAdSfr}
  f(r)=\frac{1}{\el^2}-\frac{2M}{r^3} \,,
\end{equation}
where $M$ is the parameter proportional to the mass density of the black hole\footnote{Mass density = $M/(4\pi G_N \el^2)$}, $\{x,y\}$ are dimensionless coordinates scaled by $\el$ and the horizon locates at $r=r_h=(2M\el^2)^{1/3}$.

For simplicity, we consider the complexity of a thermal state at $t_R=t_L=0$. Because of symmetry, the maximal surface is just the $t=0$ slice in the bulk. The volume of this slice is
\begin{equation}\label{VBH1}
  V=2\Omega_2\int_{r_h}^{r_m}\frac{r}{\sqrt{f(r)}}\td r \,.
\end{equation}
Here $\Omega_2$ is the area of 2-dimensional surface spanned by $x,y$ and  $r_m\rightarrow\infty$ is the UV cut-off. As a dimensionless cut-off, we introduce $\delta=\el/r_m$.
When $\delta\rightarrow0$, we can find the following expansion for integration \eqref{VBH1}
\begin{equation}\label{expandCV}
   V=\frac{\Omega_2\el^3}{\delta^2}+\frac{\Omega_2\sqrt{\pi}\Gamma(4/3)}{2\Gamma(5/6)}\el r_h^2+\mathcal{O}(\delta) \,.
\end{equation}
In fact, as shown in Ref.~\cite{Carmi:2016wjl}, such a leading divergent structure is universal and determined by the vacuum UV boundary. The effects of matter fields can affect only the finite term. It seems that we can just directly discard this divergence and use the finite term to study the effects of matter fields. If we do so, we obtain a finite result~\footnote{{From here on we set $G_N =1$.}}
\begin{equation}\label{CFAdS1}
  \mathcal{C}_{V,\text{finite}}=\frac{\el}{\ell}\frac{\Omega_2\sqrt{\pi}\Gamma(4/3)}{2\Gamma(5/6)} r_h^2\,.
\end{equation}
The process from Eq.~\eqref{expandCV} to Eq.~\eqref{CFAdS1} can be defined as a kind of background subtraction, i.e.
\begin{equation}\label{CFAdS1b}
  \mathcal{C}_{V,\text{finite}}=\frac1{\ell}\lim_{\delta\rightarrow0}\left(V-\frac{\Omega_2\el^3}{\delta^2}\right).
\end{equation}
A similar method was applied in Refs.~\cite{Albash:2012pd,Albash:2010mv,Ling:2015dma} to find the finite term of the entanglement entropy.

However, the metric \eqref{SAdSmetric1} is not the only form of the Schwarzschild AdS black hole. For example, we may use a new coordinate $\{t,r',\theta,\phi\}$ by the following coordinate transformation
\begin{equation}\label{coordtransf1}
  r=r'\left[1+\frac{F(M)\el^2}{r'^2}+F(M)\mathcal{O}(\el^4/{r'}^4)\right]^{-1}\,,
\end{equation}
when $r\gg\el$. Here $F(M)$ is an arbitrary function and $F(0)=0$. We see that, from the AdS/CFT viewpoint, there is not any physical difference between coordinate $\{t,{r'},\theta,\phi\}$  and $\{t,r,\theta,\phi\}$. Because time $t$ is not  changed, the $t=0$ slices in both coordinate are the same surface, which means their volumes, as the geometry qualities, are independent of the choice of coordinate. Let ${\delta}'=\el/{r'}_m$ be the UV cut-off in a new coordinate system. The coordinate transformation \eqref{coordtransf1} implies the following relationship between $\delta$ and ${\delta}'$
\begin{equation}\label{twodelta}
  \delta={\delta}'(1+F(M){\delta}'^2+\cdots)\,.
\end{equation}
In a new coordinate system, the volume of $t=0$ slice reads
\begin{equation}\label{drecSAdS3}
  V =\frac{\Omega_2\el^3}{{\delta}'^2}-2\Omega_2\el^3F(M)+\frac{\Omega_2\sqrt{\pi}\Gamma(4/3)}{2\Gamma(5/6)}\el r_h^2+\mathcal{O}({\delta}') \,.
\end{equation}
As expected, the leading divergent term is just as the same as Eq.~\eqref{expandCV}. However, the finite term is different!
Now assume we don't know the result in the coordinate $\{t,r,\theta,\phi\}$ and use coordinate $\{t,{r'},\theta,\phi\}$  first, then by the Eq.~\eqref{CFAdS1b}, we find that
\begin{equation}\label{complefor2}
  \mathcal{C}_{V,\text{finite}}'=  \frac{\el}{\ell} \frac{\Omega_2\sqrt{\pi}\Gamma(4/3)}{2\Gamma(5/6)} r_h^2 -2\Omega_2\el^3F(M)/\ell \,.
\end{equation}
Because $F(M)$ is arbitray, we see that a naive ``background subtraction'' yields an arbitrary regularized complexity depending on $F(M)$. Even with the geometry of the same $M$, different choices of $F(M)$ can give all different
complexity.


In recent papers \cite{Carmi:2016wjl,Reynolds:2016rvl}, the authors analysed the divergent structure of the complexity in the CV and CA conjecture in the Fefferman-Graham (FG) coordinate and, in our example case, the first term of \eqref{expandCV} is shown as a divergent term.  Naively, this divergent term can be discarded to regularize the complexity.  However, if we use another coordinate system such as \eqref{coordtransf1} different from the FG coordinate, we have to discard not only the divergent term but also a finite piece, the second term of \eqref{drecSAdS3}. Therefore, it will be better if we can identify the divergence structure of the complexity in a coordinate independent way, and subtract it to regulate the complexity. Another advantage of this coordinate independent regularized complexity lies in the computation of the complexity of formation. Unlike Ref. \cite{Chapman:2016hwi} we do not need to worry about the coordinate dependence of the cut-off.

To propose a well defined subtraction for the regularized complexity, we follow the procedure of the holographic renormalization \cite{Emparan:1999pm,deHaro:2000vlm,Skenderis:2002wp,Bianchi:2001kw}.   In this procedure, the divergences are canceled by adding covariant local boundary surface counterterms determined by the near-boundary behaviour of bulk fields.
Inspired by  \cite{Carmi:2016wjl,Reynolds:2016rvl} we use the counterterms expressed in terms of intrinsic and extrinsic curvatures.
We will show that both for the CA conjecture and CV conjecture, we can add suitable covariant local boundary counterterms to cancel the divergences appearing in the complexity. For a resolution of the example in this section see Eqs. \eqref{boundmetricexmp1}-\eqref{tt2} in section \ref{sec222}.

\subsection{Surface counterterms in CA and CV conjectures}
\subsubsection{Surface counterterms in CA conjecture}
In this subsection, we will first consider the CA conjecture. For the CA conjecture, we need to compute the action for the WDW patch. Since it has null boundaries one needs to consider appropriate boundary terms.  It was proposed in Refs.~\cite{Parattu:2015gga,Lehner:2016vdi,Jubb:2016qzt,Reynolds:2016rvl} as
\begin{equation}\label{actionull}
\begin{split}
  I=&\frac1{16\pi}\int_{\mathcal{M}}\td^{d+1}x\sqrt{-g}\left[\mathfrak{R}+\frac{d(d-1)}{\el^2}\right] \\
   &+\frac1{8\pi}\int_{\mathcal{B}}\td^dx\sqrt{|h|}\mathcal{K}-\frac1{8\pi}\int_{\mathcal{N}}\td^{d-1}x\td\lambda\sqrt{\gamma}\kappa+I_\lambda\\
  &+ \frac1{8\pi}\int_{\mathcal{J}}\td^{d-1}x\sqrt{\sigma}\eta+\frac1{8\pi}\int_{\mathcal{J'}}\td^{d-1}x\sqrt{\sigma}a  \,.
  \end{split}
\end{equation}
where the first line is the Einstein-Hilbert action with the cosmological constant integrated over the WDW region denoted by $\mathcal{M}$, the second line is various boundary terms defined at the boundary of $\mathcal{M}$ and third line is the joint terms defined on the corners of two different boundaries.
$\mathcal{B}$ stands for the time-like or space-like boundary, $\mathcal{N}$ for the null boundary,
$\mathcal{J}$ for the joints connecting time-like or space-like boundaries and $\mathcal{J}'$ for the joints connecting boundaries, one or both of which are null surfaces.   $\mathcal{K}$ is the Gibbons-Hawking-York extrinsic curvature and $h$ is the determinant of the induced metric. $\lambda$ is a parameter of the generator of the null boundary and {$\kappa$ is the non-affinity parameter of null normal vector $k^I=(\partial/\partial\lambda)^I$}, i.e., $k^I\nabla_I k^J=\kappa k^J$.  $\gamma$ is the determinant of the metric on the cross section of constant $\lambda$ in null surface $\mathcal{N}$. $\sigma$ is the induced metric at the joints. The expression for $\eta$ and $a$ can be found in Ref.  \cite{Lehner:2016vdi}.  As the joint terms $\mathcal{J}$ does not occur for the WDW patches, we will not show $\eta$ here.  $a$ is written as
{
\begin{equation}\label{expressa}
  a=\left\{
  \begin{split}
  &\pm\ln(|n^I k_I|)\,,\\
  &\pm\ln(|k^I\bar{k}_I|/2)\,,
  \end{split}
  \right.
\end{equation}
}
where $n^I$ is the unit normal vector (outward/future directed) for non-null intersecting boundary, and $\bar{k}^I$ is the other null normal vector (future directed) for null intersecting boundary. The sign in the Eq.~\eqref{expressa} can be appointed as follows: ``+'' appears only when the WDW patch appears in the future/past of null boundary component and the joint is at the past/future end of null component. It was pointed by Ref.~\cite{Lehner:2016vdi} that the action \eqref{actionull}, in its form without $I_\lambda$, depends on the parametrization of null generators. It first appeared in Ref.~\cite{Lehner:2016vdi} and was studied further in Refs.~\cite{Chapman:2016hwi,Carmi:2016wjl,Jubb:2016qzt}. Moreover, we will see later that the divergent terms in this form cannot be canceled by adding covariant surface terms.  \kyr{Thus, to make the action with the null boundaries to be invariant under the reparametrization on the null normal vector field},\footnote{For the joint terms and boundary terms, there is still an ambiguity: we may add any term of which variation vanishes.  Because the variational principle does not determine the boundary term uniquely we have a freedom to add any non-dynamic term to the complexity without any physical effects. However, if a boundary term is added at the null boundaries or the joints which are not at the AdS boundary, it may lead some dynamic effects. The physical meaning of this kind of additional freedom is not clear for us.} an additional boundary term($I_\lambda$) at the null boundaries  is added~\cite{Reynolds:2016rvl}:
%
\begin{equation}\label{addnulbd}
  I_\lambda=\mp\frac1{8\pi}\int_{\mathcal{N}}\sqrt{\gamma}\Theta\ln(\Theta\el)\td\lambda\td^{d-1}x \,,
\end{equation}
where $-$(+) appears if $\mathcal{N}$ lies to the future (past) of $\mathcal{M}$ and
\begin{equation}
\Theta = \frac{1}{\sqrt{\gamma}} \frac{\partial \sqrt{\gamma}}{\partial \lambda} \,,
\end{equation}
%


Now let us analyze how to add the surface terms so that we can obtain a finite complexity. The goal here is very similar to the case that we add some boundary terms to make the total free energy finite in holographic renormalization. However, there is an important difference. Our goal here is to make the complexity itself finite, so the surface terms do not need to be invariant under the metric variation. This admits that the surface terms can contain not only the intrinsic geometry but also the extrinsic geometry.

In the Fefferman-Graham (FG) coordinate system \cite{Graham:1999pm,2007arXiv0710.0919F}, any asymptotic AdS$_{d+1}$ space-time can be written as\footnote{{We introduce the dimensionless coordinate $z, x^\mu$ scaled by $\el$ so $\tilde{g}_{\mu\nu}$ is dimensionless and $g_{\mu\nu}$ \eqref{gmn99} has length dimension 2. All tilde-variables in this subsection are dimensionless.  }}
\begin{equation}\label{FGmetric}
  \td s^2=g_{IJ}\td x^I\td x^J=\frac{\el^2}{z^2}[\td z^2+\tilde{g}_{\mu\nu}(z,x^\mu)\td x^\mu\td x^\nu] \,,
\end{equation}
where the indices $I,J=0, 1, 2, \cdots, d-1, d$ denote the full sppace-time coordinates,  $\mu,\nu=0,2,\cdots d-1$ denote the coordinate labeled at the fixed $z$ surface.  We consider the case in which the metric $\tilde{g}_{\mu\nu}$ along the boundary directions has a power series expansion with respective to $z$ when $z\rightarrow0$:
\begin{equation}\label{powerz1}
\begin{split}
  \tilde{g}_{\mu\nu}(z,x^\mu)=\tilde{g}_{\mu\nu}^{(0)}(x^\mu)+z^2\tilde{g}_{\mu\nu}^{(1)}(x^\mu)+\cdots+z^{d}\tilde{g}_{\mu\nu}^{(d/2)}(x^\mu)+z^{d}\tilde{h}_{\mu\nu}(x^\mu)\ln z + \cdots \,,
  \end{split}
\end{equation}
where the coefficient of logarithmic term is nonzero only if $d$ is even.
\kyr{In fact, the expansion structure and coefficients of Eq. \eqref{powerz1} may be deformed by a relevant operator  (see Ref.~\cite{Hung:2011ta} for example), which will not be considered in this paper for simplicity.}
The expansion coefficients $ \tilde{g}_{\mu\nu}^{(n)} $ with $n<d/2$ and $\tilde{h}_{\mu\nu}$  are
completely determined by $\tilde{g}_{\mu\nu}^{(0)}$. The higher order coefficients are not fixed by $\tilde{g}_{\mu\nu}^{(0)}$ alone and they encode information of the expectation value of the boundary energy-momentum tensor~\cite{deHaro:2000vlm,Skenderis:2002wp}. We will see that these higher order terms are irrelevant in determining the counterterms.

At the UV cut-off $z=\epsilon$, the induced metric (denoted by $g_{\mu\nu}$)  at the boundary (codimension-one) surface is
\begin{equation} \label{gmn99}
g_{\mu\nu}=\frac{\el^2}{z^{2}}\tilde{g}_{\mu\nu} \,,
\end{equation}
and we use ``$\ \tilde{~} \ $'' to denote the conformal boundary metric at the surface $z=\epsilon$.
Likewise, in this paper, the notation ``$\tilde{X}$'' (indices are suppressed) means that it is computed by the conformal metric $\tilde{g}_{\mu\nu}$ and we use $\tilde{g}_{\mu\nu}$ to raise and lower its indexes. For example, we will  decompose the metric $g_{\mu\nu}$ as
\begin{equation}\label{boundmetric}
  g_{\mu\nu}\td x^\mu \td x^\nu=-N^2(z,t,y^i)\td t^2+\sigma_{ij}(z,t,y^i)(\td y^i-L^i\td t)(\td y^j-L^j\td t) \,,
\end{equation}
where the indices  $i,j=1,2,\cdots d-1$ and $\{x^\mu\}=\{t,y^i\}$ and we may introduce `tilde'-variables
\begin{equation}\label{boundmetriccomp}
  \tilde{N}^2=\frac{z^2}{\el^2}N^2\,, \qquad \tilde{\sigma}_{ij}=\frac{z^2}{\el^2}\sigma_{ij}\,, \qquad\tilde{L}^i=L^i\,,
\end{equation}
so
\begin{equation}\label{boundmetric}
\begin{split}
  g_{\mu\nu}\td x^\mu \td x^\nu &=\frac{\el^2}{z^2}[-\tilde{N}^2(z,t,y^i)\td t^2+\tilde{\sigma}_{ij}(z,t,y^i)(\td y^i-\tilde{L}^i\td t)(\td y^j-\tilde{L}^j\td t)]\\
  &= \frac{\el^2}{z^{2}}\tilde{g}_{\mu\nu}\td x^\mu \td x^\nu   \,.
  \end{split}
\end{equation}
Furthermore, the expansion for $\tilde{g}_{\mu\nu}$ \eqref{powerz1} can give similar expansions for $\tilde{N}$, $\tilde{\sigma}_{ij}$, and $\tilde{L}^i$:
\begin{equation}\label{expNsigma}
\begin{split}
  \tilde{N}&=\tilde{N}^{(0)}+z^2\tilde{N}^{(1)}+\cdots+z^{2[d/2]}\tilde{N}^{([d/2])}+\cdots \,, \\
   \tilde{\sigma}_{ij}&=\tilde{\sigma}_{ij}^{(0)}+z^2\tilde{\sigma}_{ij}^{(1)}+\cdots+z^{2[d/2]}\tilde{\sigma}_{ij}^{([d/2])}+\cdots \,,\\
    \tilde{L}^i&=\tilde{L}^{i(0)}+z^2\tilde{L}^{i(1)}+\cdots+z^{2[d/2]}\tilde{L}^{i([d/2])}+\cdots \,,
  \end{split}
\end{equation}
where we can fix $\tilde{N}^{(0)}=1, \tilde{L}^{i(0)}=0$ and we  can also define that
\begin{equation}\label{expNsigma2}
  \tilde{N}^{(n)}=\frac{z}{\el}{N}^{(n)}\,, \qquad \tilde{\sigma}_{ij}^{(n)}=\frac{z^2}{\el^2}{\sigma}_{ij}^{(n)} \,.
\end{equation}
As another convention, in this paper, we will always use the notation $X^{(n)}$ to denote the coefficient of $z^{(2n)}$ in the expansion of the field $X$.

Let us  consider the Ricci tensor $R_{\mu\nu}$ and the Ricci scalar $R$ for boundary metric $g_{\mu\nu}$ and the extrinsic curvature tensor $K_{ij}$ for the $t=0$ surface\footnote{Here we set $t=0$ just for convenience, we can set $t$ to be any fixed value. } (codimension-two) embedded in the $z=\epsilon$ boundary surface (codimension-one). Then we find that the conformal Ricci tensor $\tilde{R}_{\mu\nu}$, Ricci scalar $\tilde{R}$ and extrinsic curvature $\tilde{K}_{ij}$ are
\begin{equation}\label{conformaR}
  \tilde{R}_{\mu\nu}=R_{\mu\nu},~~~\tilde{R}=\frac{\el^2}{z^2}R,~~~\tilde{K}_{ij}=\frac{\el}{z}K_{ij}.
\end{equation}
For later use, we define two projections from $z=\epsilon$ surface to the $z=\epsilon$ and $t=0$ surface by ${h_i}^\mu = {{\tilde{h}}_i}^{~\mu} = \frac{\partial x^\mu}{\partial y^i}$. For example, the projections of the Ricci tensors are defined as
\begin{equation}\label{ProjrctRicci}
  \hat{R}_{ij}={h_i}^\mu{h_i}^\nu R_{\mu\nu}\,,~~~~\tilde{\hat{R}}_{ij}={{\tilde{h}}_i}^{~\nu}{{\tilde{h}}_i}^{~\nu}\tilde{R}_{\mu\nu}\,.
\end{equation}
Like the metric, we can also expand the Ricci tensor and the extrinsic curvature and other geometrical quantities with respective to $z$.

Next, we will show that the divergent terms in the action \eqref{actionull} at a given time $t$ can be reorganized as the following surface integrals
\begin{equation}\label{CAdivn}
   I_{\text{ct}}
   =  \int_B\td^{d-1}x \ \sqrt{\sigma}\,  \sum_{n=0}^{[\frac{d-1}2]}  \el^{2n}\ F_{A}^{(2n)}(d, R_{\mu\nu}, g_{\mu\nu}, \sigma_{ij},K_{ij} ) \,, 
\end{equation}
where $B$ is the codimension-two surface at a given time $t$ and fixed $z=\epsilon$.  $F_{A}^{(2n)}$ is the invariant combinations of $R_{\mu\nu}, g_{\mu\nu}, \sigma_{ij}$ and $K_{ij}$.  \kyr{The maximum level of divergence of $F^{(2n)}_A$ is $1/\epsilon^{d-1-2n}$ but $F^{(2n)}_A$ may also include less divergent terms than $1/\epsilon^{d-1-2n}$. (It is explained below Eq. \eqref{CAct1}.)}
{When the bulk dimension is even ($d$ is odd) a logarithmic divergence appears, and $F_A^{d-1}$  should be understood as a counterterm for the logarithmic divergence.}
We can define the regularized finite action, $I_{\text{reg}}$, as
\begin{equation}\label{Ireg}
  I_{\text{reg}} \equiv \lim_{\epsilon\rightarrow0}(I_\epsilon-I_{\text{ct,L}}-I_{\text{ct,R}}) \,.
\end{equation}
where $I_\epsilon$ is the action \eqref{actionull} computed with the AdS boundary at the cut-off surface $z=\epsilon$. $I_{\text{ct,L}}$ and $I_{\text{ct,R}}$ are the surface counterterms defined by \eqref{CAdivn} at the left boundary and right boundary, respectively.

\begin{figure}
  \centering
  \includegraphics[width=.46\textwidth]{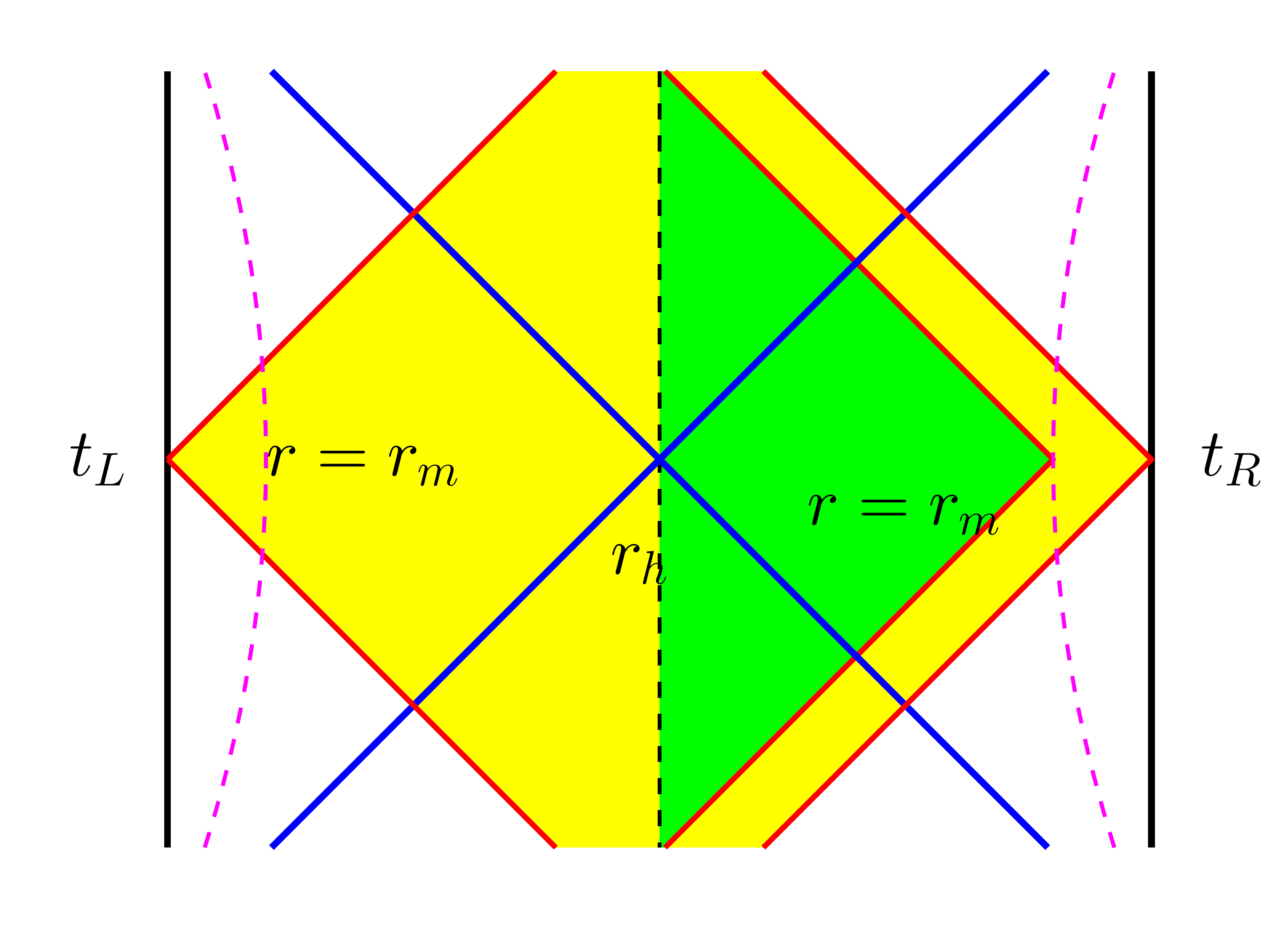}
   \includegraphics[width=.46\textwidth]{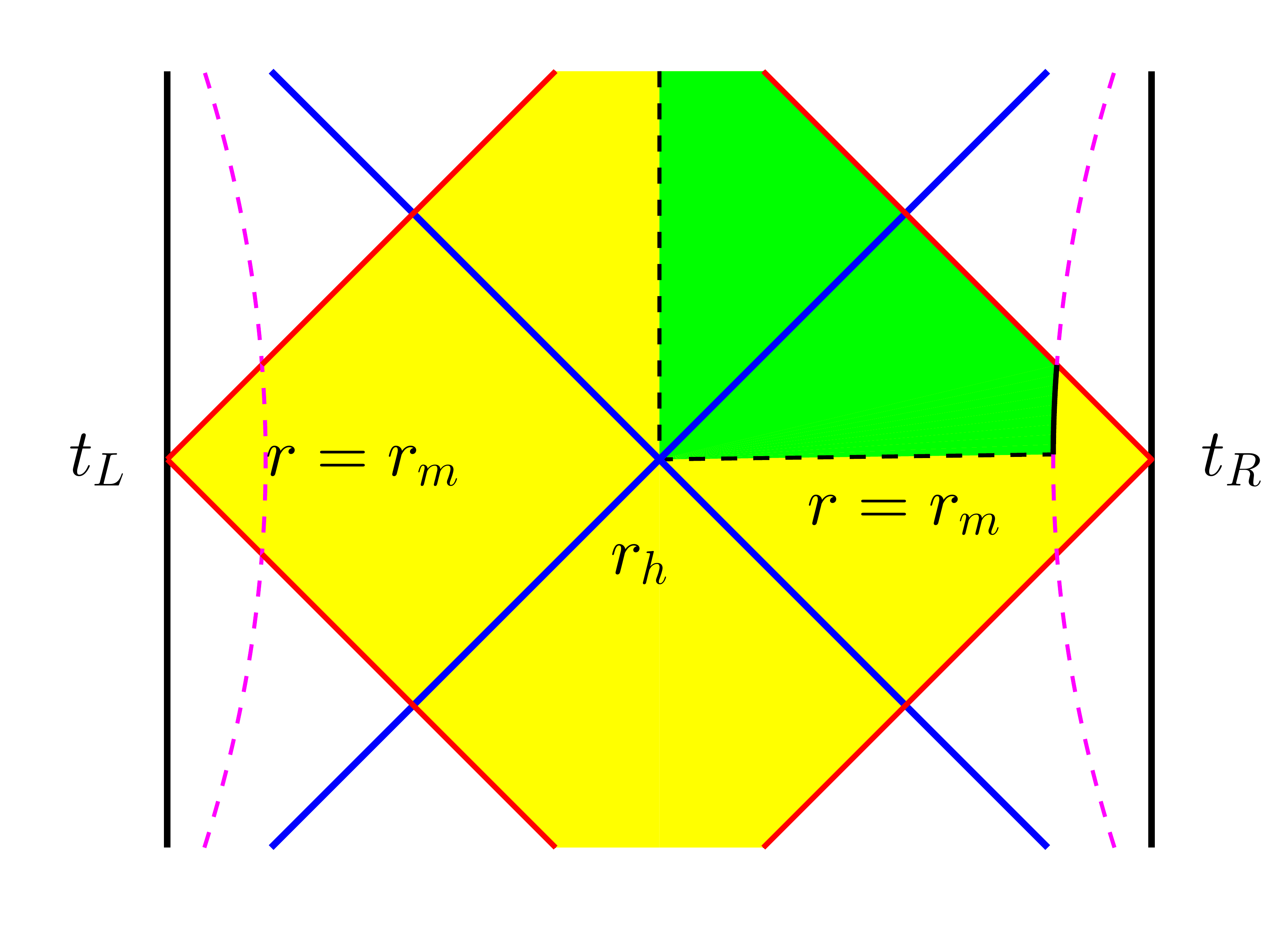}
  \caption{Two different approaches in computing the action at the finite cut-off boundaries. Left panel: The null boundaries of the WDW patch are changed into the null sheets coming from the finite cut-off boundary and there is a null-null joint at the cut-off. (here we only show the part near $t_R$. The part near $t_L$ is similar.) Right panel: The boundaries of the WDW patch are the same, but, the original null-null joints at the AdS boundary are sliced out by a time like boundary and two null-timelike joints are added. (here we only show the right-top part of quarter. The other parts are similar.)}\label{Figcutoff}
\end{figure}

Before discussing the surface counterterm $I_{\text{reg}}$  let us first explain how to compute $I_\epsilon$.  It needs to
be regulated. As pointed by Ref.~\cite{Carmi:2016wjl}, there are two different methods to regulate the WDW patch as we show in  Fig.~\ref{Figcutoff}. At the left panel of Fig.~\ref{Figcutoff}, the boundaries of the WDW patch are changed into the null sheets coming from the finite cut-off boundary and there is a null-null joint at the cut-off. At the right panel of Fig.~\ref{Figcutoff}, the boundaries of WDW patch is the same, however, original null-null joints at the AdS boundary is sliced out by a time like boundary, so the null-null joint at the boundary disappears but there is an additional Gibbos-Hawking-York boundary term and two  null-timlike joints. As the first  approach is more convenient in analyzing the divergent behavior near the AdS boundary, the term $I_\epsilon$ in the Eq.~\eqref{Ireg} is computed by this approach.

To find $I_{\text{ct}}$ or $F_{A}^{(2n)}$, we first need to analyze the divergent structure of \eqref{actionull}. The divergences come from the action near the boundary. We only need to analyze the divergent behavior at the one side boundary since the other side is similar. We will analyze the divergent behavior in the FG coordinate and show that the divergent term (the whole divergent term rather than only the coefficients of divergent terms) in this coordinate can be written as the codimension-two surface terms. After subtracting this codimension-two surface terms, we end up with the finite result. As the subtraction terms are written in terms of the geometrical quantities of the codimension-two surface, the final result is independent of the choice of coordinate. This means the result of Eq.~\eqref{Ireg} is the same for all the coordinate systems.

We first consider the case that $d+1$ is odd number. In this case there is no anomaly divergent term. All the divergent terms in the FG coordinate system are in the form of the power series of the cut-off. At any side of the two boundaries, the first two divergent terms in the action  \eqref{actionull}  were obtained in Ref.~\cite{Reynolds:2016rvl}:
\begin{equation}
I_{\text{div}} = I^{(1)}_{\text{CA}} + I^{(2)}_{\text{CA}} + \mathcal{O}\left(\frac{1}{\epsilon^{d-5}}\right) \,,
\end{equation}
where
\begin{equation}\label{CAdiv1}
  I^{(1)}_{\text{CA}} = \frac{\el^{d-1}\ln(d-1)}{4\pi\epsilon^{d-1}}\int_{B}\td^{d-1}x\sqrt{\tilde{\sigma}^{(0)}} \,,
\end{equation}
for $d\ge 2$ and\footnote{{This is different from the results reported in the Refs.~\cite{Carmi:2016wjl,Reynolds:2016rvl}. It seems that the null normal vectors used in Refs.~\cite{Carmi:2016wjl,Reynolds:2016rvl} are not affinely parameterized. If we take this into account we find an additional contribution to the subleading divergent terms.  See the appendix~\ref{app1} for details.}}
\begin{equation}\label{CAdiv2}
\begin{split}
  I^{(2)}_{\text{CA}}&=\frac{\el^{d-1}}{4\pi\epsilon^{d-3}}\int_B\td^{d-1}x\sqrt{\tilde\sigma^{(0)}}\left[\frac{\ln(d-1)}{2(d-2)}(\tilde{R}^{(0)}/2-\tilde{\hat{R}}^{(0)})\right.\\
  &\left.-\frac{d\tilde{K}^{(0)2}+2(d-1)\tilde{K}_{ij}^{(0)}\tilde{K}^{(0)ij}-3(d-1)\tilde{R}^{(0)}+2(d-1)\tilde{\hat{R}}^{(0)}}{2(d-1)(d-2)(d-3)}\right] \,,
  \end{split}
\end{equation}
for $d \ge 4$.

First, let us consider the leading divergent term $I^{(1)}_{\text{CA}}$. It is expressed in terms of the `tilde' variables \eqref{expNsigma2} and can be rewritten in terms of real induced metric as
\begin{equation}\label{CAdiv11}
  I^{(1)}_{\text{CA}}=\frac{\ln(d-1)}{4\pi}\int_B\td^{d-1}x\sqrt{\sigma^{(0)}} \,.
\end{equation}
By the inversion of the expansion to fields \eqref{expNsigma}
\begin{equation}\label{sigma1order}
  \sqrt{\sigma^{(0)}}=\sqrt{\sigma}\left(1-\frac{\epsilon^2}2\sigma^{ij}\sigma^{(1)}_{ij}\right)+\mathcal{O}(\epsilon^{5-d}) \,,
\end{equation}
we find
\begin{equation}\label{CAdiv111}
  I^{(1)}_{\text{CA}}=\frac{\ln(d-1)}{4\pi}\int_B\td^{d-1}x\sqrt{\sigma}+\mathcal{O}(\epsilon^{3-d})\,,
\end{equation}
so
\begin{equation}\label{FA1}
  F_{A}^{(0)}=\frac{\ln(d-1)}{4\pi} \,.
\end{equation}

Similarly, the subleading divergent term $I^{(2)}_{\text{CA}}$ can be rewritten as
\begin{equation}\label{CAdiv2bb}
\begin{split}
  I^{(2)}_{\text{CA}}&=\frac{\el^2}{4\pi}\int_B\td^{d-1}x\sqrt{\sigma}\left[\frac{\ln(d-1)}{2(d-2)}(R/2-\hat{R})\right.\\
  &\left.-\frac{dK^2+2(d-1)K_{ij}K^{ij}-3(d-1)R+2(d-1)\hat{R}}{2(d-1)(d-2)(d-3)}\right]+\mathcal{O}(\epsilon^{5-d}).
  \end{split}
\end{equation}
However, the first order counterterm based on Eq.~\eqref{FA1} also has contribution on the subleading divergent term. By the relationship \eqref{sigma1order} and the Einstein equations for the conformal metric $\tilde{g}_{\mu\nu}$ at the order of $\epsilon^2$ \cite{deHaro:2000vlm,Skenderis:2002wp}
\begin{equation}\label{firstEeq}
  \tilde{g}^{(1)}_{\mu\nu}=-\frac{1}{d-2}\left(\tilde{R}^{(0)}_{\mu\nu}-\frac{\tilde{g}^{(0)}_{\mu\nu}}{2(d-1)}R^{(0)}\right),
\end{equation}
we find that the subleading divergent term in the first counterterm is
\begin{equation}\label{CVdivct1}
  -\frac{\el^2}{4\pi}\int_B\td^{d-1}x\sqrt{\sigma}\frac{\ln(d-1)}{2(d-2)}(R/2-\hat{R})+\mathcal{O}(\epsilon^{5-d})\,.
\end{equation}
As a result, \eqref{CAdiv2bb} and \eqref{CVdivct1} together give the new subleading divergent term
\begin{equation}\label{CAdiv2b}
-\frac{\el^2}{4\pi}\int_B\td^{d-1}x\sqrt{\sigma}\left[\frac{dK^2+2(d-1)K_{ij}K^{ij}-3(d-1)R+2(d-1)\hat{R}}{2(d-1)(d-2)(d-3)}\right]+\mathcal{O}(\epsilon^{5-d}) \,.
\end{equation}
In other words, because the counter term \eqref{CAdiv111} already cancels the part of the subleading divergence (the first term in \eqref{CAdiv2}), we only need to introduce the second line of \eqref{CAdiv2bb} as a new counterterm.
Therefore, the second function $F_{A,2}$ is
\begin{equation}\label{expFA2}
  F_{A}^{(2)}=-\frac{1}{4\pi}\frac{dK^2+2(d-1)K_{ij}K^{ij}-3(d-1)R+2(d-1)\hat{R}}{2(d-1)(d-2)(d-3)} \,.
\end{equation}
%


\kyr{The general structure of divergences in the CA conjecture was suggested to be~\cite{Carmi:2016wjl}
\begin{equation}\label{generdivCA}
  I_{\text{div}}=  \frac{\el^{d-1}}{G_N} \int_{B}\td^{d-1}x\frac{\sqrt{\tilde{\sigma}^{(0)}}}{\epsilon^{d-1}}\sum_{n=0}^{[\frac{d-1}2]}\epsilon^{2n} \sum_{i}\tilde{c}_{i,n}(d)[ \tilde{\mathcal{R}}^{(0)}, \tilde{K}^{(0)} ]_{i}^{2n}\,.
\end{equation}
where we dropped the log term in  Ref.~\cite{Carmi:2016wjl} by considering  the null boundary term \eqref{addnulbd} following  Ref.~\cite{Reynolds:2016rvl}. We recovered $G_N$ here to make clear $I_{\text{div}}$ is dimensionless.
$[ \tilde{\mathcal{R}}^{(0)}, \tilde{K}^{(0)} ]_{i}^{2n}$ is a schematic expression indicating invariant   combinations of  $\tilde{R}_{\mu\nu}^{(0)}, \tilde{K}_{ij}^{(0)}$, $\tilde{g}_{\mu\nu}^{(0)}$ and $\tilde\sigma^{(0)}_{ij}$ with a mass dimension of $2n$.  Thus, $\epsilon^{2n}[ \tilde{\mathcal{R}}^{(0)}, \tilde{K}^{(0)} ]_{i}^{2n}$ is dimensionless.  The index $i$ stands for a different combination. For example, in Eq.~\eqref{CAdiv1} there is only one term (say $i=1$) and one can read $[ \tilde{\mathcal{R}}^{(0)}, \tilde{K}^{(0)} ]_{1}^{0} = 1$ with $\tilde{c}_{1,0}=\ln(d-1)/(4\pi)$. In Eq.~\eqref{CAdiv2} there are four invariant combinations $[ \tilde{\mathcal{R}}^{(0)}, \tilde{K}^{(0)} ]_{i}^{2} $=$\{ \tilde{K}^{(0)2}, \tilde{K}_{ij}^{(0)} \tilde{K}^{(0)ij},$ $\tilde{R}^{(0)}, \tilde{\hat{R}}^{(0)}\}$ with the corresponding coefficients $\tilde{c}_{i,1}$ which can  be read from Eq.~\eqref{CAdiv2}. \kyb{To be more concrete, $\tilde{c}_{i,1}$ was summarized in Table. \ref{tab:1}. } For some symmetry arguments for the divergence structure and pattern, we refer to Ref.~\cite{Carmi:2016wjl}.}

\begin{table}
\centering
\begin{tabular}{|l|l|}
\hline
$\tilde{c}_{1,0} = \frac{\ln (d-1)}{4\pi} $&  ${c}_{1,0} = \frac{\ln (d-1)}{4\pi} $    \\
\hline
$  \tilde{c}_{1,1} =  -\frac{d}{8\pi(d-1)(d-2)(d-3)}\quad $  & $  {c}_{1,1} =-\frac{d}{8\pi(d-1)(d-2)(d-3)}  \qquad$ \\
$  \tilde{c}_{2,1} = - \frac{1}{4\pi(d-2)(d-3)}\quad $  & $  {c}_{2,1} =- \frac{1}{4\pi(d-2)(d-3)}\qquad$ \\
$  \tilde{c}_{3,1} = \frac{\ln (d-1)}{16\pi (d-2)} + \frac{3}{8\pi(d-2)(d-3)} \quad  $  & $  {c}_{3,1} = \frac{3}{8\pi(d-2)(d-3)} \qquad$ \\
$  \tilde{c}_{4,1} = -\frac{\ln (d-1)}{8\pi (d-2)} - \frac{1}{4\pi(d-2)(d-3)}\quad $  & $  {c}_{4,1} = - \frac{1}{4\pi(d-2)(d-3)} \qquad$ \\
\hline
\end{tabular}
\caption{$\tilde{c}_{1,0}$ and $c_{1,0}$ are read from Eq.~\eqref{CAdiv1} and Eq.~\eqref{CAdiv111}.  $\tilde{c}_{i,1}$ is the coefficient of $[ \tilde{\mathcal{R}}^{(0)}, \tilde{K}^{(0)} ]_{i}^{2} $=$\{ \tilde{K}^{(0)2}, \tilde{K}_{ij}^{(0)} \tilde{K}^{(0)ij},\tilde{R}^{(0)}, \tilde{\hat{R}}^{(0)}\}$ in Eq.~\eqref{CAdiv2} and $c_{i,1}$ is the coefficient of $[ {\mathcal{R}}, {K} ]_{i}^{2} $=$\{ {K}^{2}, {K}_{ij} {K}^{ij},{R}, {\hat{R}}\}$ in Eq.~\eqref{CAdiv2b}. It is valid up to holographic spacetime dimension 5 or less. } \label{tab:1}
\end{table}

 \kyr{Once we obtain the divergent structures \eqref{generdivCA} for $d \ge 5$, we can repeat the steps that we have done for $d=3,4$. Thus we propose that the following counter terms work for $d \ge 5$ as well as $d=3,4$.
\begin{equation} \label{CAct1}
  I_{\text{ct}}= \frac{\el^{d-1}}{G_N} \int_{B}\td^{d-1}x \frac{\sqrt{{\sigma}}}{\el^{d-1}}\sum_{n=0}^{[\frac{d-1}2]}\el^{2n}\sum_{i}c_{i,n}(d)[\mathcal{R}, K]_i^{2n}\,.
\end{equation}
This is similar to Eq.~\eqref{generdivCA} in structure. $\epsilon$ in Eq.~\eqref{generdivCA} is absorbed to $\tilde{\sigma}^{(0)}$ and $\tilde{\mathcal{R}}^{(0)}$, leaving $\el$ to take into account dimension.  However, note that the structure of $[ \tilde{\mathcal{R}}^{(0)}, \tilde{K}^{(0)} ]_{i}^{2n}$ and $[\mathcal{R}, K]_i^{2n}$ are not the same as shown in Eq.~\eqref{CAdiv2} and Eq.~\eqref{CAdiv2b}.  i.e. the expected level of divergence of $[\mathcal{R}, K]_i^{2n}$ is equal to or less than $1/\epsilon^{d-1-2n}$. \kyb{To be more concrete, ${c}_{i,1}$ was summarized in Table. \ref{tab:1}.}
Finally, for notational convenience, we rewrite \eqref{CAct1}  as, with $G_N=1$,
\begin{align}
  I_{\text{ct}}
   = &  \int_B\td^{d-1}x \ \sqrt{\sigma}\,  \sum_{n=0}^{[\frac{d-1}2]}  \el^{2n}\ F_{A}^{(2n)}(d, R_{\mu\nu}, g_{\mu\nu}, \sigma_{ij},K_{ij} )\,, \\ 
& \mathrm{with} \ \ F_{A}^{(2n)}(d, R_{\mu\nu}, g_{\mu\nu}, \sigma_{ij},K_{ij}) \equiv  \sum_{i} c_{i,n}(d)[\mathcal{R}, K]_i^{2n} \,,
\end{align}
which is \eqref{CAdivn}. }

\kyb{
In order to show the explicit formulas for higher dimensions than 5, we first have to obtain the divergence structure explicitly similar to Eq \eqref{CAdiv2} following Refs.~\cite{Carmi:2016wjl,Reynolds:2016rvl}. We think it can be done straightforwardly but the final formulas will be very complicated. Thus,  it may not be so illuminating for the purpose of explaining the methodology.
However, it is possible  to obtain the simple formulas for some special cases. This case is explained in detail in appendix~\ref{app2}.}


\kyr{For all the cases that $d$ is odd integer greater than 1,  there is a logarithmic divergent term\footnote{One should note that if we don't add $I_\lambda$ into the action  \eqref{actionull}, the additional logarithm divergent term will appear \cite{Carmi:2016wjl} in any dimension. In general, it has following forms
\begin{equation}\label{logdivCA}
  I_{\log,\text{CA}}=\ln(\sqrt{\alpha\beta}\epsilon/\el)\left(\frac{c_1}{\epsilon^{d-1}}+\frac{c_2}{\epsilon^{d-3}}\cdots\right).
\end{equation}
Here coefficients $c_1, c_2,\cdots$ are determined by the conformal boundary metric $\tilde{g}_{\mu\nu}^{(0)}$ but $\alpha, \beta$ are arbitrary constants depending on the choice of null normal vectors in null surfaces. As the $\alpha$ and $\beta$ can not be determined by theory itself, such terms cannot be written as the covariant geometrical quantities of the  boundary metric. This results show that it is necessary to add the term $I_\lambda$ into the action \eqref{actionull} to obtain an covariant regularized complexity.
}  in the action \eqref{actionull}. At the cut-off $\epsilon$ in any given coordinate system, the counterterm is given by
\begin{equation}\label{CAlog1}
  -\ln(\epsilon/\el)\int_B\td^{d-1}x\sqrt{\sigma}\el^{d-1}\bar{F}_{A}^{(d-1)}+\mathcal{O}(\epsilon)\,.
\end{equation}
Here $\bar{F}_{A}^{(d-1)}=\lim_{d'\rightarrow d}(d'-d)F_{A}^{(d-1)}$. Note the integration term in Eq.~\eqref{CAlog1} is finite and coordinate-independent. }

After we obtain the regularized form of the action in the WDW region, we propose to define a ``regularized complexity''  as follows
\begin{equation}\label{newCA}
  \mathcal{C}_{\text{A,reg}}=\lim_{\epsilon\rightarrow0}\frac1{\pi\hbar}(I_\epsilon-I_{\text{ct,L}}-I_{\text{ct,R}})=\frac{I_{\text{reg}}}{\pi\hbar}\,.
\end{equation}
This is similar to the holographic renormalization of the on-shell action for a free energy.
In the holographic renormalization of the free energy \cite{Emparan:1999pm}, the counterterms are only intrinsic geometric quantities not to affect the equations of motion. However, when we regularize the complexity, this restriction may be relaxed and the extrinsic quantities may be included.
For both a free energy and the complexity, the relative value between two states is important so a subtraction of the same value from two states are allowed.  The complexity describes the minimum number of quantum gates required to produce some state  from a particular reference state, so it does not matter if we add any constant value in complexity to  both states.  As a reference state we can appoint any non-dynamic quantum state. The subtraction term $I_{\text{ct,R}}$ and $I_{\text{ct,R}}$ are defined by the boundary metric and does not contain any bulk dynamics and matter fields information, so they are non-dynamic subtraction terms.  Therefore, we can consider $ \mathcal{C}_{\text{A,reg}}$  as a well defined ``regularized compelxity'' in the CA conjecture.
%

\subsubsection{Surface counterterms in CV conjecture} \label{sec222}
Similarly to the CA case,  we can define the regularized complexity for the CV conjecture as
\begin{equation}\label{newCV}
   \mathcal{C}_{\text{V,reg}}=\lim_{\epsilon\rightarrow0}\frac1{\ell}(V_\epsilon-V_{\text{ct,L}}-V_{\text{ct,R}}) \,,
\end{equation}
where $V_\epsilon$ is the maximum value connecting $t_L$ and $t_R$ after we use a finite cut-off $z=\epsilon$ to replace the real AdS boundary.   $V_{\text{ct,L}}$ and $V_{\text{ct,R}}$ are the surface counterterms
\begin{equation}\label{expCVF}
  V_{\text{ct}}=\el \int_B\td^{d-1}x \ \sqrt{\sigma}\,  \sum_{n=0}^{[\frac{d-1}2]}  \el^{2n}\ F_{V}^{(2n)}(d, R_{\mu\nu}, g_{\mu\nu}, \sigma_{ij},K_{ij} )\,, 
\end{equation}
at the left boundary ($V_{\text{ct,L}}$) and right boundary ($V_{\text{ct,R}}$) respectively.
When the bulk dimension is even ($d$ is odd) a logarithmic divergence appears, and $F_{V}^{d-1}$  should be understood as a counterterm for the logarithmic divergence.

We first consider the odd bulk dimensions. To find $V_{\text{ct}}$ or $F_{V,n}$ we first need to analyze the divergent structure of \eqref{CV}. The first two divergent terms in the volume \eqref{CV} for $d \ge 2$ were obtained in Ref.~\cite{Reynolds:2016rvl}:
\begin{equation}
V_{\text{div}} = V^{(1)} + V^{(2)} + \mathcal{O}\left(\frac{1}{\epsilon^{d-5}}\right) \,,
\end{equation}
where
\begin{equation} \label{tt1}
V^{(1)} =\frac{\el^{d}}{d-1}\int_B\td^{d-1}x\sqrt{\tilde{\sigma}^{(0)}}\epsilon^{1-d} \,,
\end{equation}
and
\begin{equation} \label{yyyy1}
V^{(2)} =-\frac{\el^{d}(d-1)}{2(d-2)(d-3)\epsilon^{d-3}}\int_B\td^{d-1}x\sqrt{\tilde{\sigma}^{(0)}}\left[\tilde{\hat{R}}^{(0)}-\frac{\tilde{R}^{(0)}}2-\frac{(d-2)^2}{(d-1)^2}\tilde{K}^{(0)2}\right]\,.
\end{equation}
With these two equations, the first volume divergent term reads
\begin{equation}\label{CVdiv1}
  V^{(1)}=\frac{\el}{d-1}\int_B\td^{d-1}x\sqrt{\sigma} +\mathcal{O}(\epsilon^{3-d})\,,
\end{equation}
so we have
\begin{equation}\label{F1CV}
  F_{V}^{(0)}=\frac1{d-1} \,.
\end{equation}
The subleading divergent term can be written as
\begin{equation}\label{CVdiv2}
  V^{(2)}=-\frac{\el^3}{2(d-2)(d-3)}\int_B\td^{d-1}x\sqrt{\sigma}\left[\hat{R}-\frac{R}2-\frac{(d-2)^2}{(d-1)^2}K^2\right]+\mathcal{O}(\epsilon^{5-d})\,.
\end{equation}
As the same as the CA conjecture, the first surface counterterm has also contribution on the subleading divergence
\begin{equation*}
\frac{\el^3}{2(d-2)(d-1)}\int_B\td^{d-1}\sqrt{\sigma}\left[\hat{R}-\frac{R}2\right]+\mathcal{O}(\epsilon^{5-d})\,,
\end{equation*}
which leads that total subleading divergent term reads
\begin{equation}\label{CVsubdiv}
  -\frac{\el^3}{2(d-2)(d-3)}\int_B\td^{d-1}x\sqrt{\sigma}\left[\frac2{d-1}(\hat{R}-R/2)-\frac{(d-2)^2}{(d-1)^2}K^2\right]+\mathcal{O}(\epsilon^{5-d}) \,,
\end{equation}
so we obtain
\begin{equation}\label{CVsubdiv2}
  F_{V}^{(2)}=-\frac{1}{2(d-2)(d-3)}\left[\frac2{d-1}(\hat{R}-R/2)-\frac{(d-2)^2}{(d-1)^2}K^2\right].
\end{equation}
Such step can be continued for higher dimensional case, so we see that we can use codimension-two surface terms as the counterterms to cancel all the divergences in the volume \eqref{CV}.\footnote{When we finished this paper, we noted two Refs.~\cite{Gover:2016xwy,Gover:2016hqd} which also developed a general regulated volume expansion for the volume of a manifold with boundary. It will be interesting to study if this is equivalent to our method when it is applied to the CV conjecture.}

\kyr{The general structure of divergences in the CV conjecture was suggested to be~\cite{Carmi:2016wjl}
\begin{equation}\label{generdivCV}
  V_{\text{div}}=\el^d  \int_{B}\td^{d-1}x\frac{\sqrt{\tilde{\sigma}^{(0)}}}{\epsilon^{d-1}}\sum_{n=0}^{[\frac{d-1}2]}\epsilon^{2n} \sum_{i}\tilde{c}_{i,n}(d)[ \tilde{\mathcal{R}}^{(0)}, \tilde{K}^{(0)} ]_{i}^{2n}\,.
\end{equation}
The structure is the same to the CA case \eqref{generdivCA} apart from the overall factor $\el^d$ accounting for the dimension of volume. However, the explicit expressions for $\tilde{c}_{i,n}$ and $[ \tilde{\mathcal{R}}^{(0)}, \tilde{K}^{(0)} ]_{i}^{2n}$ are different from the CA case.
$[ \tilde{\mathcal{R}}^{(0)}, \tilde{K}^{(0)} ]_{i}^{2n}$ is a schematic expression indicating invariant   combinations of  $\tilde{R}_{\mu\nu}^{(0)}, \tilde{K}_{ij}^{(0)}$, $\tilde{g}_{\mu\nu}^{(0)}$ and $\tilde\sigma^{(0)}_{ij}$ with a mass dimension of $2n$.  Thus, $\epsilon^{2n}[ \tilde{\mathcal{R}}^{(0)}, \tilde{K}^{(0)} ]_{i}^{2n}$ is dimensionless.  The index $i$ stands for a different combination. For example, in Eq.~\eqref{tt1} there is only one term (say $i=1$) and one can read $[ \tilde{\mathcal{R}}^{(0)}, \tilde{K}^{(0)} ]_{1}^{0} = 1$ with $\tilde{c}_{1,0}=1/(d-1)$. In Eq.~\eqref{yyyy1} there are three invariant combinations $[ \tilde{\mathcal{R}}^{(0)}, \tilde{K}^{(0)} ]_{i}^{2} $=$\{\tilde{\hat{R}}^{(0)}, \tilde{R}^{(0)},  \tilde{K}^{(0)2}\}$ with the corresponding coefficients $\tilde{c}_{i,1}$ which can  be read from Eq.~\eqref{yyyy1}. \kyb{To be more concrete, $\tilde{c}_{i,1}$ was summarized in Table. \ref{tab:2}. } For some symmetry arguments for the divergence structure and pattern, we refer to Ref.~\cite{Carmi:2016wjl}.}

\begin{table}
\centering
\begin{tabular}{|l|l|}
\hline
$\tilde{c}_{1,0} = \frac{1}{d-1} $&  ${c}_{1,0} = \frac{1}{d-1} $    \\
\hline
$  \tilde{c}_{1,1} = -\frac{d-1}{2(d-2)(d-3)} \qquad $  & $ {c}_{1,1}= -\frac{1}{(d-1)(d-2)(d-3)} \quad $ \\
$  \tilde{c}_{2,1} = \frac{d-1}{4(d-2)(d-3)} \qquad $  & $  {c}_{2,1} = \frac{1}{2(d-1)(d-2)(d-3)} \quad$ \\
$  \tilde{c}_{3,1} =  \frac{d-2}{2(d-1)(d-3)} \qquad $  & $  {c}_{3,1} = \frac{d-2}{2(d-1)^2(d-3)} \quad  $ \\
\hline
\end{tabular}
\caption{$\tilde{c}_{1,0}$ and $c_{1,0}$ are read from Eq.~\eqref{tt1} and Eq.~\eqref{CVdiv1}.  $\tilde{c}_{i,1}$ is the coefficient of $[ \tilde{\mathcal{R}}^{(0)}, \tilde{K}^{(0)} ]_{i}^{2} $=$\{ \tilde{\hat{R}}^{(0)} ,\tilde{R}^{(0)},\tilde{K}^{(0)2} \}$ in Eq.~\eqref{yyyy1} and $c_{i,1}$ is the coefficient of $[ {\mathcal{R}}, {K} ]_{i}^{2} $=$\{ {\hat{R}, {R}, {K}^{2} }\}$ in Eq.~\eqref{CVsubdiv}. It is valid up to holographic spacetime dimension 5 or less.} \label{tab:2}
\end{table}

\kyr{Once we obtain the divergent structures \eqref{generdivCV} for $d \ge 5$, we can repeat the steps that we have done for $d=3,4$. Thus we propose that the following counter terms work for $d \ge 5$ as well as $d=3,4$.
\begin{equation} \label{yyyy2}
  V_{\text{ct}}=\el^d\int_{B}\td^{d-1}x \frac{\sqrt{{\sigma}}}{\el^{d-1}}\sum_{n=0}^{[\frac{d-1}2]}\el^{2n}\sum_{i}c_{i,n}(d)[\mathcal{R}, K]_i^{2n}\,,
\end{equation}
This is similar to Eq.~\eqref{generdivCV} in structure. $\epsilon$ in Eq.~\eqref{generdivCV} is absorbed to $\tilde{\sigma}^{(0)}$ and $\tilde{\mathcal{R}}^{(0)}$, leaving $\el$ to take into account dimension.  However, note that the structure of $[ \tilde{\mathcal{R}}^{(0)}, \tilde{K}^{(0)} ]_{i}^{2n}$ and $[\mathcal{R}, K]_i^{2n}$ are not the same as shown in Eq.~\eqref{yyyy1} and Eq.~\eqref{CVsubdiv}.  i.e. the expected level of divergence of $[\mathcal{R}, K]_i^{2n}$ is equal to or less than $1/\epsilon^{d-1-2n}$. \kyb{To be more concrete, ${c}_{i,1}$ was summarized in Table. \ref{tab:2}. }
Finally, for notational convenience, we rewrite \eqref{yyyy2}  as
\begin{align}
  V_{\text{ct}}&=\el \int_B\td^{d-1}x \ \sqrt{\sigma}\,  \sum_{n=0}^{[\frac{d-1}2]}  \el^{2n}\ F_{V}^{(2n)}(d, R_{\mu\nu}, g_{\mu\nu}, \sigma_{ij},K_{ij} )\,, \\ 
   & \mathrm{with} \ \  F_{V}^{(2n)}(d, R_{\mu\nu}, g_{\mu\nu}, \sigma_{ij},K_{ij}) \equiv  \sum_{i} c_{i,n}(d)[\mathcal{R}, K]_i^{2n}
\end{align}
which is \eqref{expCVF}. }
\kyb{
To find the explicit formulas for higher dimensions than 5, we first have to obtain the divergence structure explicitly similar to Eq. \eqref{yyyy1} following Refs.~\cite{Carmi:2016wjl,Reynolds:2016rvl}. We think it can be done straightforwardly but the final formulas will not be so illuminating. However, similarly to the CA case,
it is possible to obtain the simple formulas for some special cases. It is shown in detail in appendix~\ref{app2}.}

\kyr{When the bulk dimension $d+1$ is even, the logarithmic divergent term will appear, which is similar to the the case in the CA conjecture. The counterterm at the cut-off $\epsilon$ in any coordinate system reads
\begin{equation}\label{addlogCV}
  \ln(\epsilon/\el)\el^d\int_B\td^{d-1}x\sqrt{\sigma}\bar{F}_{V}^{(d-1)}+\mathcal{O}(\epsilon)\,.
\end{equation}
Here $\bar{F}_{V}^{(d-1)}=\lim_{d'\rightarrow d}(d'-d){F}_{V}^{(d-1)}$. Note that the integration in the equation is  finite and coordinate independent. }

As an example, let us compute the regularized complexity by the CV conjecture for the example shown in section \ref{exmp1}. The metric of the boundary and the codimension-two surface at $t=0$ are
\begin{equation}\label{boundmetricexmp1}
  g_{\mu\nu}\td x^\mu\td x^\nu=-r^2f(r)\td t^2+r^2(\td x^2+\td y^2)\,, \qquad \sigma_{ij}\td x^i\td x^j=r^2(\td x^2+\td y^2).
\end{equation}
The Ricci tensor is zero at the boundary at fixed $r=\el/\delta$ and the extrinsic curvature is also zero at the surface of $t=0$ embedding in the boundary.  So the subleading term in Eq. \eqref{CVsubdiv2} is zero and there is only one term in the surface counterterm, which reads
\begin{equation}
  V_{\text{ct,L}}=V_{\text{ct,R}}=\frac{\el}2\int\td^2x\sqrt{\sigma}=\frac{\Omega_2\el^3}{2\delta^2}.
\end{equation}
We see that this is just the value shown in Eq. \eqref{VBH1}. So in this coordinate system, the surface counterterm is as the same as the background subtraction term and we find the regularized complexity is just as the same as one shown in Eq. \eqref{CFAdS1}. Of course, we can also compute the regularized complexity in the coordinate $\{t,r',x,y\}$, where the relationship between $r$ and $r'$ is given by Eq. \eqref{coordtransf1}. The surface counterterm at the cut-off ${\delta}'$ then is
\begin{equation} \label{tt2}
  V_{\text{ct,L}}=V_{\text{ct,R}}=\frac{\Omega_2\el^3}{2\delta'^2}-\Omega_2\el^3F(M)+\mathcal{O}(\delta').
\end{equation}
We see that in this coordinate system, the counterterm is not proportional to the volume of the pure AdS space-time, as its value depends on mass $M$. However, one can find that the regularized complexity is still as the same as Eq. \eqref{CFAdS1}, which is independent of the choice of $F(M)$.

We want to stress that  it is important to use \eqref{CVdiv1} as a subtraction term rather than \eqref{tt1}.  If we used \eqref{tt1} as a subtraction term, we would not have \eqref{tt2} so the regularized complexity becomes coordinate dependent and ambiguous. In sections.~\ref{exmplBTZ} and~\ref{exmplSAdS}, we will give more examples for computing the regularized complexity for the CV and CA conjectures.

Note also that our surface counterterms are non-dynamic and have no relationship to the bulk matter field, so such subtraction keeps all the information of bulk matter field in the complexity. In addition, if there is asymptotic time-like Killing vector field $\xi=(\partial/\partial t)^\mu$ at the boundary  we have
\begin{equation}\label{timederivative}
  \frac{\td \mathcal{C}_{\text{reg}}}{\td t}=\frac{\td \mathcal{C}}{\td t}\,.
\end{equation}
This means the previous studies about the complexity growth, in fact, studied the behavior of the regularized part of the whole complexity.

If we let $\gamma$ be any parameter in the system which has no effect on the boundary metric, and we can define a ``complexity of formation'' between two different states labeled by $\gamma=\gamma_1$ and $\gamma=\gamma_2$ as
\begin{equation}\label{Coffmt}
  \Delta\mathcal{C}=\mathcal{C}_{\text{reg}}(\gamma_1)-\mathcal{C}_{\text{reg}}(\gamma_2)\,.
\end{equation}
If $\gamma$ is the temperature and $\gamma_1=T, \gamma_2=0$,  \eqref{Coffmt} gives the complexity of formation studied in Ref.~\cite{Chapman:2016hwi}.

\section{Examples for BTZ black holes}\label{exmplBTZ}
In this section, we will give examples to compute the regularized complexity for both CA and CV conjectures in the BTZ black holes. One form of the metric for the rotational BTZ black hole is  \cite{Banados:1992wn,Banados:1992gq},
\begin{equation}\label{metricBTZ}
  \td s^2=-r^2f(r)\td t^2+\frac{\td r^2}{r^2f(r)}+r^2\left(\td\varphi-\frac{J\td t}{2r^2}\right)^2 \,,
\end{equation}
with $r\in(0,\infty)$, $\varphi\in[0,2\pi]$ and the function $f(r)$ is described by
\begin{equation}\label{BTZfr}
  f(r)=\frac1{\el^2}-\frac{M}{r^2}+\frac{J^2}{4r^4}=\frac1{\el^2}\left[1-(r_+/r)^2\right]\left[1-(r_-/r)^2\right] \,,
\end{equation}
where $M$ is the mass parameter\footnote{The physical mass for the BTZ black hole is $M/8$.} and $J$ is the angular momentum:
\begin{equation}\label{MJr1r2}
  M=\frac{r_+^2+r_-^2}{\el^2},~~~J=\frac{2r_-r_+}{\el}\,.
\end{equation}
This black hole arises from the identifications of points of the anti-de Sitter space by a discrete subgroup of SO(2, 2). The surface $r = 0$ is not a curvature singularity but, rather, a singularity in the causal structure if $J\neq0$. Although the  parameter $M$ plays the role of mass, it is possible to admit $M$ to be negative when $J=0$. In these cases, except for $M=-1$, naked conical singularities appear, so these cases should be prohibited. In the special case that $J=0$ and $M=-1$, the conical singularity disappears. The configuration is just the pure AdS$_3$ solution with $f(r)=r^2/\el^2+1$. For the case that $J>0$, we need that $M\geq J$ to avoid the naked singularity.

The BTZ black hole also has thermodynamic properties similar to those found in higher dimensions. We can define the temperature $T$, entropy $S$ and {angular velocity} $\Omega$ as
\begin{equation}\label{BTZTS}
  T=\frac{r_+^2-r_-^2}{2\pi r_+},~~~S=\frac{\pi r_+}{2},~~~\Omega=\frac{r_-}{r_+\el}\,.
\end{equation}


\subsection{CA conjecture in non-rotational case}
\begin{figure}
  \centering
  \includegraphics[width=.46\textwidth]{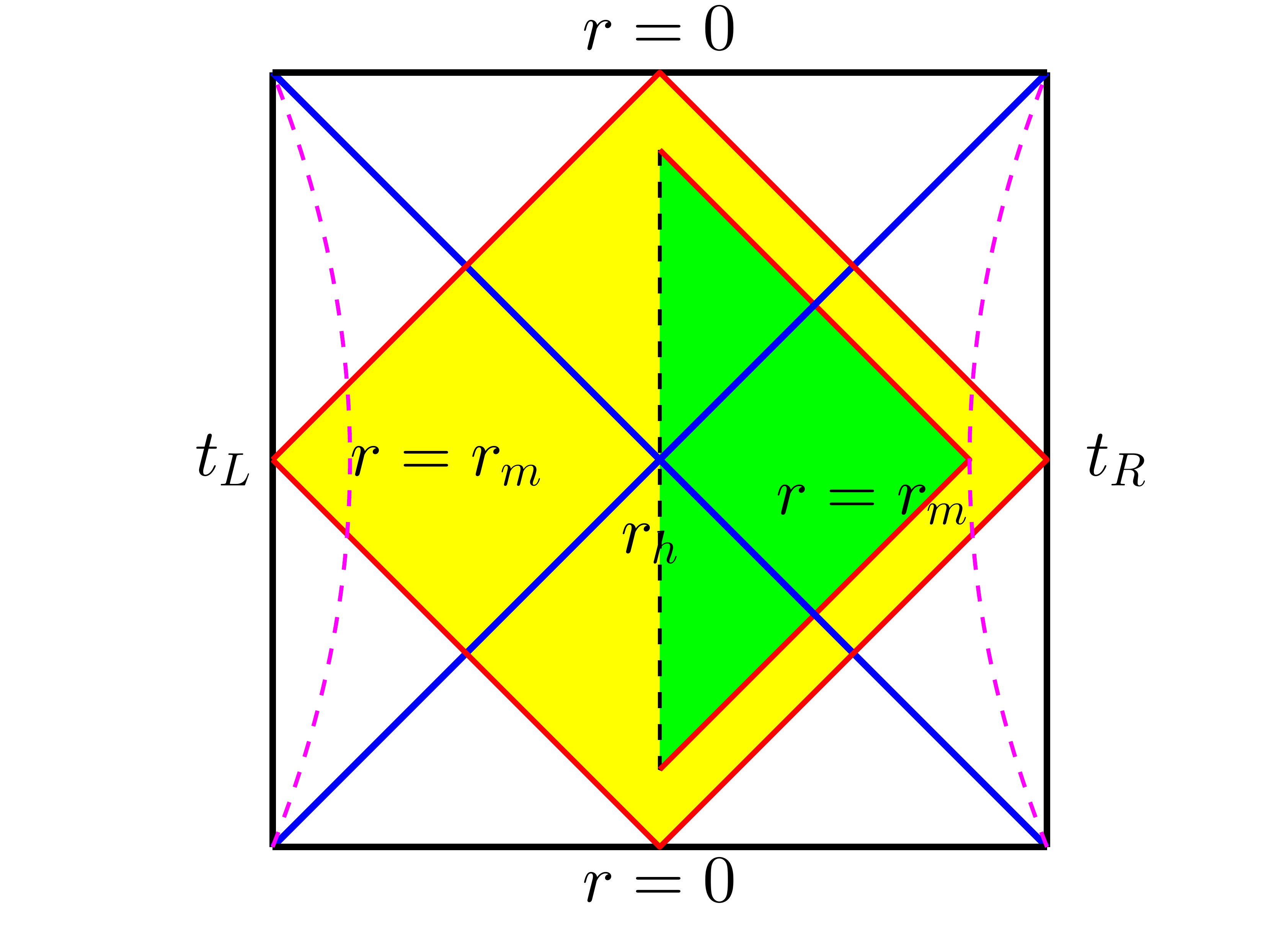}
   \includegraphics[width=.46\textwidth]{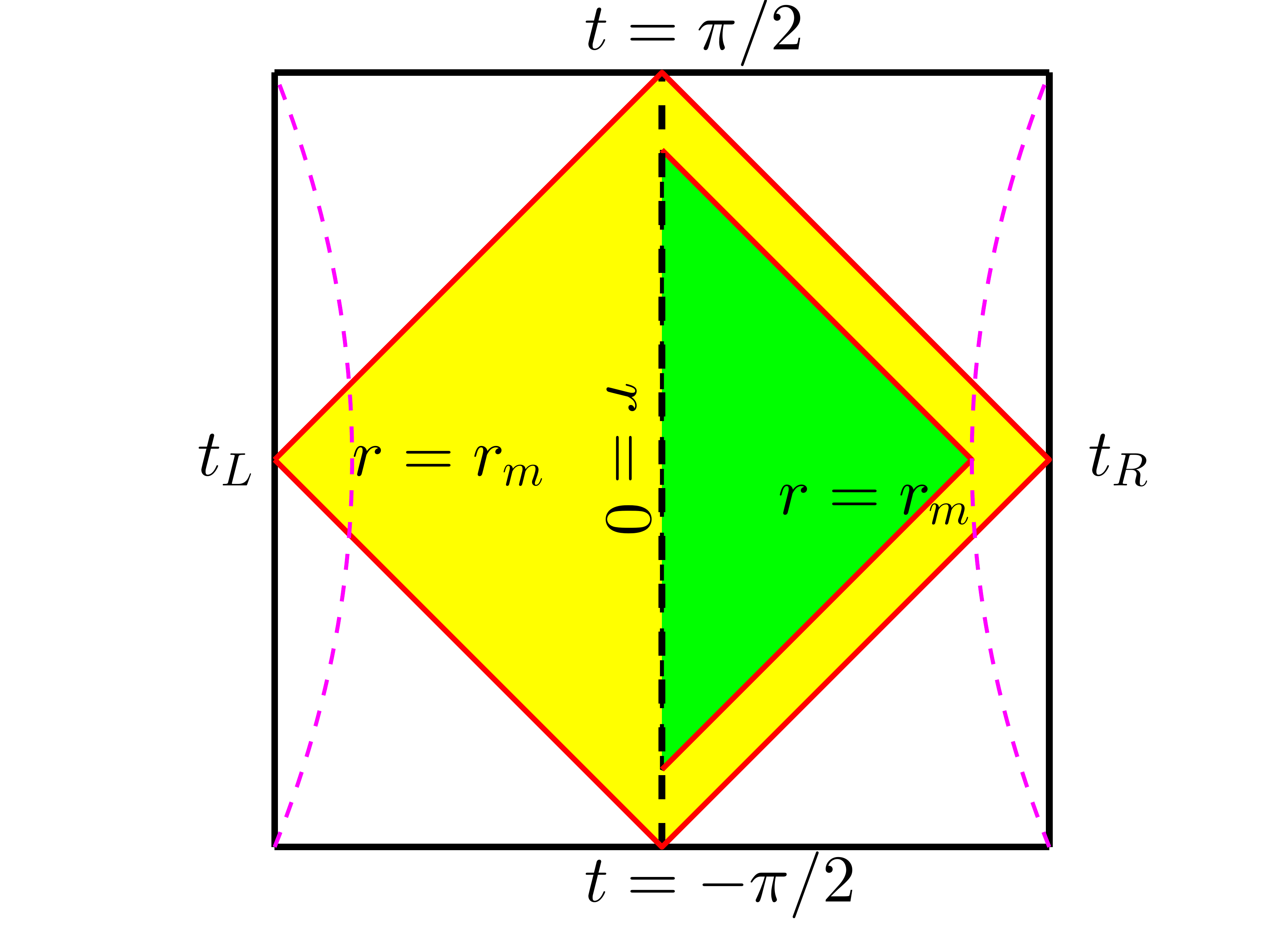}
  \caption{Penrose diagram and the regularized WDW patch for $t_R=t_L=0$ in the BTZ black holes when $J=0$. The case for $M\geq0$ is shown in the left panel, where the null sheets coming from $t_R$ and $t_L$ meet each other at the surface $r=0$. The case for $M=-1$ is shown in the right panel, where the null sheets coming from $t_R$ and $t_L$ will meet each other at $r=0$ and $t=\pm\pi/2$. }\label{FigBTZ}
\end{figure}

We first consider the case that $J=0$.  For the case $M>0$ the WDW patch is shown in the left panel of Fig.~\ref{FigBTZ}. We define $r_h \equiv r_+=\el\sqrt{M}$. For a special case of $t_L=t_R=0$, the null sheets coming from left boundary and right boundary just meet with each other at $r=0$.

In order to compute the regularized complexity in the CA conjecture, we first need to regularize the WDW patch, which is shown in the left panel of Fig.~\ref{FigBTZ}. Note that this approach is different from the approach in Ref. \cite{Chapman:2016hwi}.  Taking the symmetry into account, we only need to compute the bulk term, the boundary terms and joints at the green region. Let us introduce the outgoing and infalling null coordinates $u,v$ defined by\footnote{Strictly speaking, this relationship for $r^*$ and $r$ can only be used when $r_h>0$. However, we can see that it has a well defined limit when $r_h\rightarrow0^+$. So the  $M=0$ case can be regarded as the limit of $r_h\rightarrow0^+$. }
\begin{equation}\label{infallingv1}
\begin{split}
  &u(t,r)=t-r^*\,, \qquad v(t,r)=t+r^*(r)\,, \\
  &r^*(r)=\int [r^2f(r)]^{-1}\td r=\frac{\el^2}{2r_h}\ln\left|\frac{r-r_h}{r+r_h}\right|+v_0 \,,
\end{split}
\end{equation}
where $v_0$ is an integration constant. The null boundaries at the green region in left panel of Fig.~\ref{FigBTZ} is given by $v=v_m$ and $u=u_m$, where $v_m \equiv v(0,r_m)=r^*(r_m)$ and $u_m\equiv u(0,r_m)=-r^*(r_m)$. One can check that the dual normal vectors for such null boundaries are $k_I=\alpha[(\td t)_I+r^{-2}f^{-1}(\td r)_I]$ and $\bar{k}_I=\beta[(\td t)_I-r^{-2}f^{-1}(\td r)_I]$. Here we explicitly exhibit the freedom of choosing dual normal vector by two arbitrary constants $\alpha$ and $\beta$.  In the green region of Fig. \ref{FigBTZ}, there are a bulk integration term, two null boundary terms, a null-null joint term in the action.

Using the method similar to Ref. \cite{Chapman:2016hwi}, the bulk action is expressed as
\begin{equation}\label{BTZbulk}
  I_{\text{bulk}}=-\frac2{\el^2}\int_0^{r_m}[v_m-r^*(r)]r\td r=-r_m+\mathcal{O}(1/r_m)\,,
\end{equation}
where the factor 2 is multiplied to take the both sides ($t_L$ and $t_R$) into account.  It is different from the result in Ref.~\cite{Chapman:2016hwi} because we used a different regularization method. However,  the final results of the complexity will be the same. As the measurement of null-null joints at the corners $r=0$ is zero, such joint term has no contribution to the action. The joint term at the boundary is given by following expression
\begin{equation}\label{BTZjoint}
  I_{\text{joint}}=\left.-\frac12r\ln(|k^I\bar{k}_I|/2)\right|_{r=r_m}=\left.\frac12r\ln\left[\frac{r^2f(r)}{\alpha\beta}\right]\right|_{r=r_m}.
\end{equation}
Since $k_{I}$ is affinely parameterized, only the null boundary term shown in Eq. \eqref{addnulbd} has contribution. The expansions of $k_I$ and $\bar{k}_I$ are
\begin{equation}\label{expnsionk}
  \Theta=g^{IJ}\nabla_I k_{J}=\frac{\alpha}r,~~\bar{\Theta}=g^{IJ}\nabla_I \bar{k}_{J}=-\frac{\beta}r.
\end{equation}
In order to compute the   value of $I_\lambda$, we need to find the affine parameter $\lambda$ and $\bar{\lambda}$ for $k^I$ and $\bar{k}^I$, respectively. On the null boundary of the green region shown in the Fig.~\eqref{FigBTZ}, the coordinates $t$ and $r$ are the functions of $\lambda$, i.e., $t=t(\lambda)$ and $r=r(\lambda)$. By the equation
\begin{equation}\label{eqlambdaBTZ}
  k^I=\left(\frac{\partial}{\partial\lambda}\right)^I=\frac{\td t}{\td\lambda}\left(\frac{\partial}{\partial t}\right)^I+\frac{\td r}{\td\lambda}\left(\frac{\partial}{\partial r}\right)^I,
\end{equation}
we see that $\lambda=r/\alpha$ for $k^I$. Similarly, we find that $\lambda=-r/\beta$ for $\bar{k}^I$. So we obtain that
\begin{equation}\label{IlambdaBTZ}
  I_\lambda=-2\cdot\frac1{8\pi}\int_0^{2\pi}\td\varphi\int_{r_m}^0\alpha^{-1}\td r r\frac{\alpha}r\ln\left(\frac{\el\alpha}r\right)+(\alpha\rightarrow\beta)=\frac{r_m}2\ln\left(\frac{\alpha\beta\el^2}{r_m^2}\right)+r_m.
\end{equation}

Adding up all results, we have
\begin{equation}\label{IctBTZ}
I_{\text{reg}}(M\geq0)=I_{\text{bulk}}+I_{\text{joint}}+I_{\lambda}=0\Rightarrow\mathcal{C}_{\text{A,reg}}(M\geq0)=0,
\end{equation}
so the regularized complexity is zero for all $M\geq0$. Note that the complexity is already finite without any regularization in this case. Indeed, for $d=2$, the counterterm we derived in \eqref{CAdiv111} is always zero so our computation here is consistent.
We also see that the regularized complexity is independent of the choice of $\alpha$ and $\beta$, which is expected as $\alpha$ and $\beta$ are gauge degrees of freedom in the choices of the dual normal vector for null surface. Note that the UV divergent behavior shown in Ref.~\cite{Carmi:2016wjl} depends on these two gauge parameters. However, in our formula, as the additional term $I_\lambda$ has been added into the action \eqref{actionull}, the final result is independent of the gauge choices on the null normal vector fields.

When $M=-1$, the expression of $r^*$ in the Eq.~\eqref{infallingv1} should be replaced by following equation
\begin{equation}\label{BTZRstar2}
  r^*=-\el\arctan(\el/r)+v_0.
\end{equation}
In this case, we see that the null sheets coming from the $r=\infty,~t=0$ will meet each other at the position of $r=0$ and $t=\mp[r^*(0)-v_0]=\pm\pi/2$ respectively (see the right panel of Fig.~\ref{FigBTZ}). The computation of the regularized complexity is very similar to the case of $M>0$. Eq.~\eqref{BTZbulk} can still be used to compute the bulk term, but the result now becomes
\begin{equation}\label{BTZbulk2}
  I_{\text{bulk}}=-r_m-\frac{\pi\el}{2}+\mathcal{O}(1/r_m).
\end{equation}
The joint term at $r=r_m$ and the null boundary term have the same expressions shown in Eq.~\eqref{BTZjoint} and \eqref{IlambdaBTZ}.
Therefore, without any counterterm
\begin{equation}\label{IctBTZ2}
I_{\text{reg}}(M=-1)=I_{\text{bulk}}+I_{\text{joint}}+I_{\lambda}=\frac{\pi\el}2\Rightarrow\mathcal{C}_{\text{A,reg}}(M=-1)=\frac{\el}{2\hbar}.
\end{equation}

Using our regularized complexity, we can compute the complexity of formation \eqref{Coffmt}:
\begin{equation}
\Delta\mathcal{C} = \mathcal{C}_{\text{reg}}(M\geq0) -  \mathcal{C}_{\text{reg}}(M=-1) = - \frac{\el}{2\hbar} \,,
\end{equation}
which reproduces the result in Ref.~\cite{Chapman:2016hwi}. Because $M=-1$ has lower energy, it is the vacuum solution rather than the case with the limit $M\rightarrow0$.

\subsection{CA conjecture in rotational case}
\begin{figure}
  \centering
  \includegraphics[width=.55\textwidth]{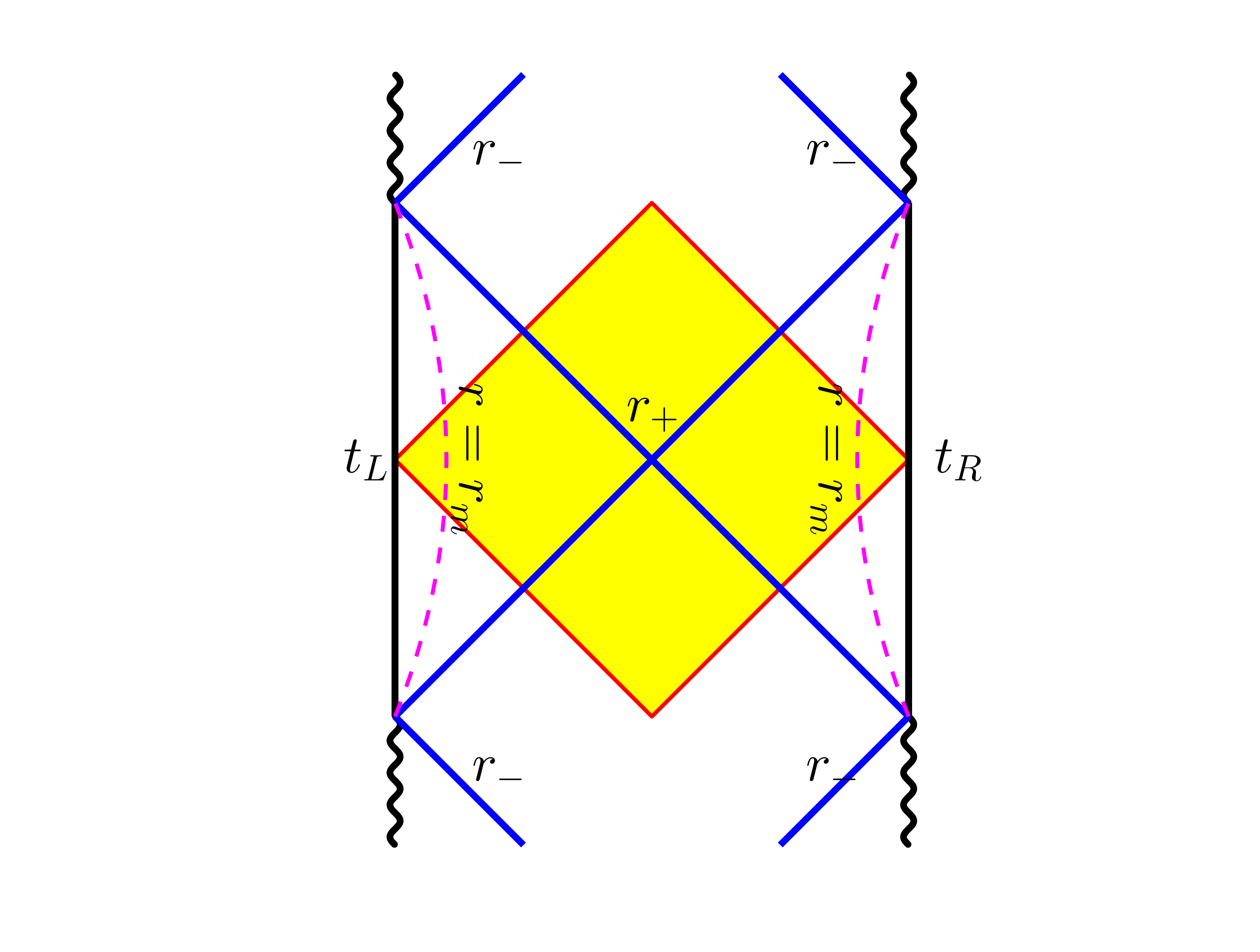}
   \includegraphics[width=.44\textwidth]{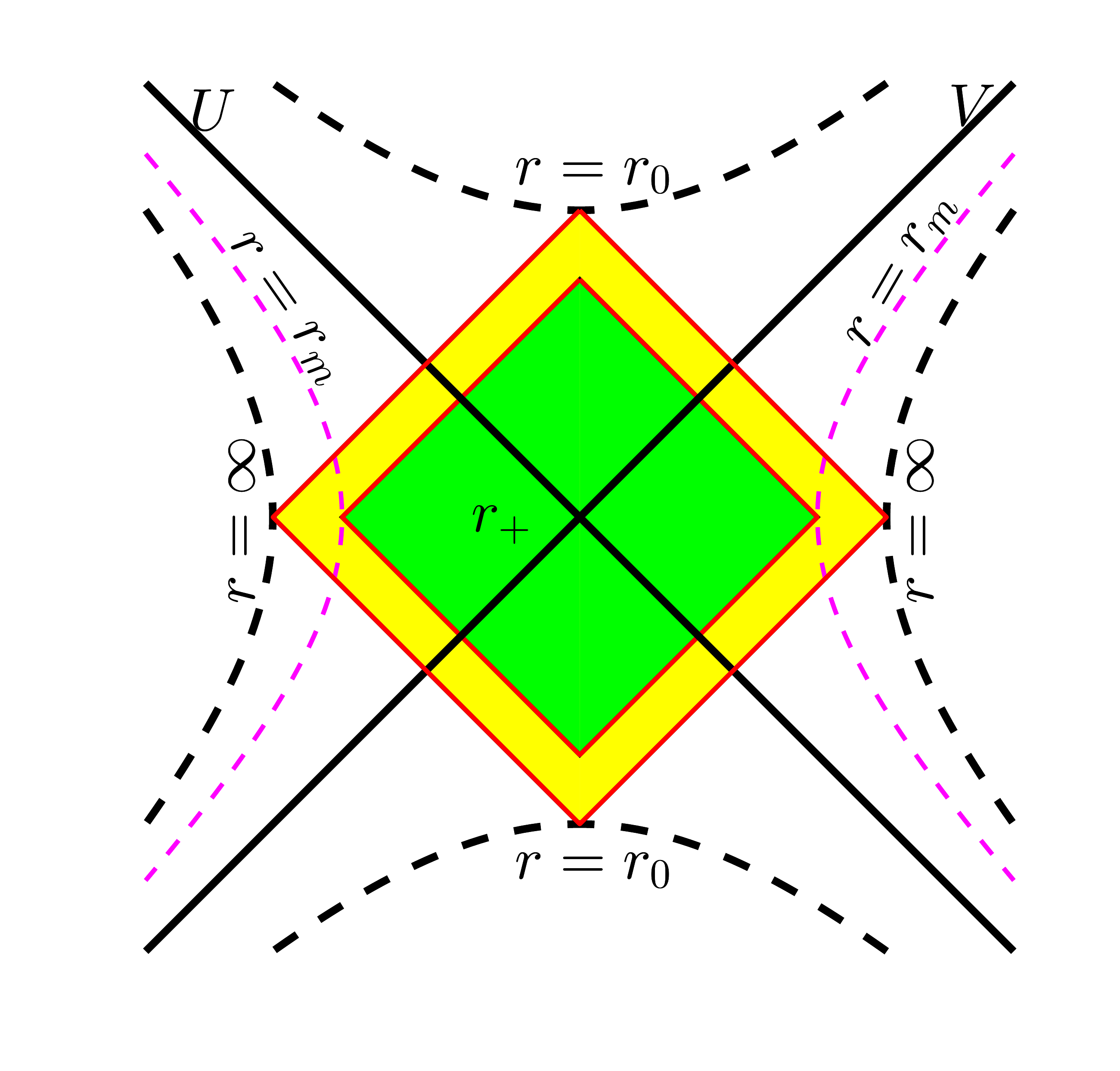}
  \caption{The WDW patch in the Penrose diagram (left panel) and the Kruskal-type coordinate (right panel) for the rotational BTZ black hole. The null sheets coming from $t_R=t_L=0$ meet each other at the surface $r=r_0\in(r_-,r_+)$.}\label{FigBTZ2}
\end{figure}

For the case that $J\neq0$, the mass $M$ must be non-negative value. There is an inner horizon behind in the outer horizon. In this case, the Penrose diagram and the WDW patch is shown in the left panel in Fig.~\ref{FigBTZ2}. As the same as the case of $J=0$, we introduce the infalling coordinate and outgoing coordinate $u$ and $v$ by the Eq.~\eqref{infallingv1}, however, the function $r^*(r)$ then becomes
\begin{equation}\label{Jrstar}
  r^*=\int\frac{\td r}{r^2f(r)}=\frac{\el^2}{2(r_+^2-r_-^2)}\left[r_-\ln\left(\frac{r+r_-}{|r-r_-|}\right)-r_+\ln\left(\frac{r+r_+}{|r-r_+|}\right)\right].
\end{equation}
In the region $r\in(r_-,\infty)$, the function $r^*(r)$ has two monotonic regions, i.e., $(r_+,\infty)$ and $(r_-, r_+)$. From Eq.~\eqref{Jrstar}, we find that $r^*(\infty)=0, r^*(r_+)=-\infty$ and $r^*(r_-)=\infty$. In this case, the null sheets coming from the $t_L$ and $t_R$ meet each other at the inner region of event horizon at finite radius $r=r_0\neq0$. The value of $r_0$ can be determined by equation $r^*(r)=0$ with the restriction $r<r_+$. Then we obtain the following transcendental equation
\begin{equation}\label{eqr0}
r_-\ln\left(\frac{r_0+r_-}{r_0-r_-}\right)-r_+\ln\left(\frac{r_0+r_+}{r_+-r_0}\right)=0.
\end{equation}

The computation for the regularized action is very similar to what we have done at the case of $J=0$. The bulk term can be computed by the same formula shown in the Eq.~\eqref{BTZbulk} but the lower limit of the integration is $r_0$, i.e.
\begin{equation}\label{BTZbulk2}
  I_{\text{bulk}}=-\frac2{\el^2}\int_{r_0}^{r_m}[v_m-r^*(r)]r\td r=-r_m+I_0(r_0)+\mathcal{O}(1/r_m),
\end{equation}
where $I_0$ is defined by
\begin{equation}\label{defineI0BTZ}
\begin{split}
  I_0(r_0)&=r_0-\frac{r_+}2\ln\left(\frac{r_0+r_+}{r_+-r_0}\right)\,.
  \end{split}
\end{equation}
The null boundary term shown in Eq.~\eqref{IlambdaBTZ} now reads
\begin{equation}\label{IlambdaBTZ2}
  I_\lambda=\frac{r_m}2\ln\left(\frac{\el^2}{r_m^2}\right)-\frac{r_0}2\ln\left(\frac{\el^2}{r_0^2}\right)+r_m-r_0.
\end{equation}
As the final result is independent on the choice of $\alpha, \beta$, we have fixed $\alpha=\beta=1$. The contribution of the joint terms at the boundary $r=r_m$ have the same formula as the $J=0$ case but  we need to add the joint term at $r=r_0$ since they have nonzero values
\begin{equation}\label{BTZjoint2}
  I_{\text{joint}}=\frac12r_m\ln[r_m^2f(r_m)]-\frac12r_0\ln[-r_0^2f(r_0)].
\end{equation}

Finally, combining the results in Eqs. \eqref{BTZbulk2}, \eqref{IlambdaBTZ2} and \eqref{BTZjoint2}, we find that
\begin{equation}\label{IctBTZ2}
\begin{split}
I_{\text{reg}}&=I_{\text{bulk}}+I_{\text{joint}}+I_{\lambda}=-\frac{r_+}2\ln\left(\frac{r_0+r_+}{r_+-r_0}\right)-\frac{r_0}2\ln\left[-\el^2f(r_0)\right]\\
&=-\frac{r_0}2\ln\left[(r_+^2-r_0^2)(r_0^2-r_-^2)/r_0^4\right]-\frac{r_+}2\ln\left(\frac{r_0+r_+}{r_+-r_0}\right).
\end{split}
\end{equation}
where there is no surface counterterms since $d=2$. This result goes to zero when $r_-\rightarrow 0$ (so $r_0 \rightarrow 0$), which reproduces the case with $J=0$ shown in \eqref{IctBTZ}.
Because $I_{\text{reg}}/r_+$ depends on only $r_-/r_+=\Omega\el$ we can introduce an auxiliary dimensionless function $\hat{I}(x)$ such that $I_{\text{reg}}/r_+ \equiv \hat{I}(\Omega\el) $. Or
\begin{equation}\label{IJBTZ}
  I_{\text{reg}}(T, \Omega)=\frac{2\pi T}{1-\Omega^2\el^2}\hat{I}(\Omega\el)=\frac{2\hat{I}(\Omega\el)}{\pi}S.
\end{equation}
where we used the expressions in \eqref{BTZTS} for $r_0$. The value of $\hat{I}(\Omega\el)$ can be computed  only numerically with $r_0$ determined by \eqref{eqr0}.

Let us consider two special cases. First, for small momentum case
($\Omega\el\ll1$)
\begin{equation}\label{IctBTZ3}
 \hat{I}(\Omega\el)=c_0\Omega\el\ln(\Omega\el)+\cdots \,,
\end{equation}
where $c_0\approx1.19967\cdots$.
Second, for low temperature case ($1-\Omega\el\ll1$),
\begin{equation}\label{IctBTZ3}
\hat{I}(\Omega\el)=-\frac12\ln(1-\Omega\el)+\cdots \,.
\end{equation}
Note that $\hat{I}(\Omega\el)$ is less than zero for small $\Omega\el$ but larger than zero for large $\Omega\el$.

Using our regularized complexity, we can compute the complexity of formation \eqref{Coffmt},
$ \Delta\mathcal{C} = \mathcal{C}_{\text{reg}} -  \mathcal{C}_{\text{reg}}(M=-1) $,
which reproduces the result in Ref.~\cite{Chapman:2016hwi}.

\subsection{CV conjecture in BTZ black hole}
Now let us calculate the regularized complexity for the CV conjecture in the BTZ black hole.
For simplicity, we consider the complexity of a thermal state defined on the time slice $t_R=t_L=0$. The maximal volume is just like Eq.~\eqref{VBH1}
\begin{equation}\label{}
V_\delta=4\pi\int_{r_+}^{r_m}\frac1{\sqrt{f(r)}}\td r=\frac{4\pi\el^2}{\delta}+4\pi\int_{r_+}^{r_m}\left(\frac1{\sqrt{f(r)}}-\el\right)\td r-4\pi\el r_+.
\end{equation}
Here we introduce a cut-off at the boundary by $r=r_m=\el/\delta$. In this case, we only need one surface term Eq.~\eqref{CVdiv1},
\begin{equation}\label{CVdiv1a}
V^{(1)}_{\text{ct}}=\el\int_B\td \varphi\sqrt{\sigma}=\frac{2\pi\el^2}{\delta}.
\end{equation}
Then the regularized complexity Eq.~\eqref{newCV} can be written as
\begin{equation}\label{CVBTZ}
\mathcal{C}_{\text{V,reg}}=\lim_{\delta\rightarrow0}\frac1{\ell}(V_\delta-2V^{(1)}_{\text{ct}})=4\pi\ell^{-1}\int_{r_+}^{\infty}\left(\frac1{\sqrt{f(r)}}-\el\right)\td r-4\pi r_+\el\ell^{-1}.
\end{equation}

Let us first consider the case  $J=0$. For $M\geq0$, it turns out that
\begin{equation}\label{CVBTZm1}
\mathcal{C}_{\text{V,reg}}=0.
\end{equation}
For $M=-1$, there is no horizon and the regularized complexity is
\begin{equation}\label{CVBTZm2}
\mathcal{C}_{\text{V,reg,vac}}=4\pi\ell^{-1}\int_{0}^{\infty}\left(\frac1{\sqrt{f(r)}}-\el\right)\td r=-4\pi\el^2\ell^{-1}\,,
\end{equation}
where the subscript ``vac'' is added since $M=-1$ is the lowest energy state. Using our regularized complexity, we can compute the complexity of formation \eqref{Coffmt}:
\begin{equation}
\Delta\mathcal{C} = \mathcal{C}_{\text{reg}}(M\geq0) -  \mathcal{C}_{\text{reg}}(M=-1) =  4\pi \el \,,
\end{equation}
which reproduces the result in Ref.~\cite{Chapman:2016hwi} if we choose $\ell = \el$.

Next, let us consider the case $J\neq0$.  The regularized complexity Eq.~\eqref{CVBTZ} yields
\begin{equation}\label{}
\frac{\ell}{\el}\mathcal{C}_{\text{V,reg}}=4\pi\int_{r_+}^{\infty}\left(\frac{r^2}{\sqrt{(r^2-r_+^2)(r^2-r_-^2)}}-1\right)\td r-4\pi r_+.
\end{equation}
Like \eqref{IctBTZ2}, by introducing $\hat{\mathcal{C}}=\ell\el^{-1}\mathcal{C}_{\text{V,reg}}/r_+$, we have
\begin{equation}\label{}
\frac{\ell}{\el}\mathcal{C}_{\text{V,reg}}(T, \Omega)=\frac{2\pi T\el^2}{1-\Omega^2\el^2}\hat{\mathcal{C}}(\Omega\el)=\frac{2\hat{\mathcal{C}}(\Omega\el)}{\pi}S \,,
\end{equation}
where the value of $\hat{\mathcal{C}}(\Omega\el)$ can be determined only numerically.

Let us consider two special cases. First, for small momentum case
($\Omega\el\ll1$)
\begin{equation}\label{}
\hat{\mathcal{C}}(\Omega\el)=\pi^2(\Omega\el)^2+\cdots \,.
\end{equation}
Second, for low temperature case ($1-\Omega\el\ll1$), we find
\begin{equation}\label{xx1}
\frac{\hat{\mathcal{C}}(\Omega\el)}{4\pi}=-\frac{1}{2}\text{ln}(1-\Omega\el)+\text{ln}(2\sqrt{2})-1+\cdots.
\end{equation}
Interestingly, low temperature behaviour of the CA and CV conjectures are similar. Indeed they are exactly the same if we choose $\ell=4\pi^2\hbar\el$.

We conclude this section by showing how to derive \eqref{xx1} in detail.
First we consider the leading behavior of $\hat{\mathcal{C}}(\Omega\el)$ at  low temperature limit, i.e., $\Omega\el\rightarrow1$. If we define $x=r/r_+$ and $x_-=r_-/r_+=\Omega\el$  the volume integral $\hat{\mathcal{C}}(\Omega\el)$ can be written as
\begin{equation}\label{yy1}
\begin{split}
\frac{\hat{\mathcal{C}}(\Omega\el)}{4\pi}&=\int_{1}^{\infty}\left(\frac{x^2}{\sqrt{(x^2-1)(x^2-x_-^2)}}-1\right)\td x-1\\
&=\int_{1}^{\infty}\left(\frac1{\sqrt{(x-1)P(x,x_-)}}-1\right)\td x-1\\
&=\left(\int_{1}^{a}\td x+\int_{a}^{\infty}\td x\right)\left(\frac1{\sqrt{(x-1)P(x,x_-)}}-1\right)-1\\
&=\int_{1}^{a}\frac{\td x}{\sqrt{(x-1)P(x,x_-)}}+ \text{finite term}
\end{split}
\end{equation}
where $0<{x_-}<1$ and $a(a>1)$ is any constant. $P(x,x_-)$ is defined as
\begin{equation}\label{}
\begin{split}
(x-1)P(x,x_-)&\equiv(x-1)(x-{x_-})(x+1)(x+{x_-})/x^4\\
&\equiv(x-1)(x-{x_-})h(x,{x_-})\,,
\end{split}
\end{equation}
where the second line defines another function $h(x,x_-)$ for convenience. To read off the singular part of Eq. \eqref{yy1} we define $H(x_-)$ as
\begin{equation}\label{}
\begin{split}
H({x_-})&\equiv\int_{1}^{a}\frac{\td x}{\sqrt{(x-1)(x-{x_-})h_0}}\\
&=\frac{2}{\sqrt{h_0}}\text{ln}(\sqrt{x-1}+\sqrt{x-{x_-}})|_1^a\\
&=-\frac1{\sqrt{h_0}}\text{ln}(1-{x_-})+\text{finite term}\,,
\end{split}
\end{equation}
where $h_0\equiv h(1,{x_-})=2(1+{x_-})$.
On the other hand, we have
\begin{equation}\label{}
\begin{split}
&\frac{\hat{\mathcal{C}}(\Omega\el)}{4\pi}-H({x_-})\\
=&\int_{1}^{a}\left(\frac{\td x}{\sqrt{(x-1)(x-{x_-})h(x,{x_-})}}-\frac{\td x}{\sqrt{(x-1)(x-{x_-})h(1,{x_-})}}\right)+ \text{finite term}\\
=&\int_{1}^{a}\frac{h_0-h(x,{x_-})}{\sqrt{(x-1)(x-{x_-})h(x,{x_-})h_0(\sqrt{h_0}+\sqrt{h(x,{x_-})})}}\td x+ \text{finite term}\\
=&\text{finite value}.
\end{split}
\end{equation}
Therefore, for the limit $\Omega \el \rightarrow 1$, we find that
\begin{equation}\label{}
\frac{\hat{\mathcal{C}}(\Omega\el)}{4\pi}=-\frac{1}{2}\text{ln}(1-\Omega\el)+\text{ln}(2\sqrt{2})-1+\cdots.
\end{equation}

\section{Examples for Schwarzschild AdS$_{d+1}$ black holes}\label{exmplSAdS}
A general Schwarzschild  AdS$_{d+1}$ ($d\geq3$) black hole is given by following metric
\begin{equation}\label{SAdSmetric}
  \td s^2=-r^2f(r)\td t^2+\frac{\td r^2}{r^2f(r)}+r^2\td\Sigma_{d-1,k}^2
\end{equation}
with
\begin{equation}\label{SAdSfr}
  f(r)=\frac{k}{r^2}+\frac{1}{\el^2}-\frac{\omega^{d-2}}{r^d} \,.
\end{equation}
Here $\omega$ is the `mass' parameter
\begin{equation}
\omega^{d-2} = r^{d-2}_h \left(  \frac{r_h^2}{\el} + k  \right) \,,
 \end{equation}
with the horizon position $r_h$ and $k=\{1,0,-1\}$ corresponding to spherical, planar and hyperbolic horizon. The $(d-1)$-dimensional line element $\td\Sigma_{d-1,k}^2$ is given by
\begin{equation}\label{Sigmak1}
  \td \Sigma_{d-1,k}^2=\left\{
  \begin{split}
  &\td\theta^2+\sin^2\theta\td\Omega_{d-2}^2,~~~~k=1;\\
  &\sum_{i=1}^{d-1}\td x^2_i,~~~~~~~~~~~k=0;\\
  &\td\theta^2+\sinh^2\theta\td\Omega_{d-2}^2,~~k=-1.
  \end{split}
  \right.
\end{equation}
Here $\Omega_{d-2}^2$ is a line element of $d-2$ dimensional unit sphere.    The dimensionless volume of the spatial geometry will be denoted by $\Sigma_{d-1,k}$.  The horizon locates at $r=r_h$. For simplicity, we still consider the case $t_R=t_L=0$ and try to find the regularized complexity in both the CA and CV conjectures. In this paper, we will only focus on the cases of $d=3,4$.

\subsection{Regularized complexity in CA conjecture}
\subsubsection*{Case of $d=3$}
In order to compute the regularized complexity in the CA conjecture, let us first introduce the outgoing and infalling null coordiantes $u,v$ defined by the same manner shown in Eq.~\eqref{infallingv1}, but the function $r^*$ now should be changed as
\begin{equation}\label{SAdSrstars}
  r^*=\frac{r_h\el^2}{3r_h^2+k\el^2}\ln\left(\frac{|r-r_h|}{\sqrt{r^2+rr_h+r_h^2+k\el^2}}\right)+\frac{2v_{\infty}}{\pi}\arctan\left(\frac{2r+r_h}{\sqrt{3r^2_h+4k\el^2}}\right)\,,
\end{equation}
with
\begin{equation}\label{vinfty}
  v_{\infty}=\frac{\pi\el^2(3r_h^2+2k\el^2)}{2(3r_h^2+k\el^2)\sqrt{3r_h^2+4k\el^2}}.
\end{equation}
In order not to make the computation too complicated, we assume first $r_h>2\el/\sqrt{3}$ when $k<0$.

\begin{figure}
  \centering
  \includegraphics[width=.49\textwidth]{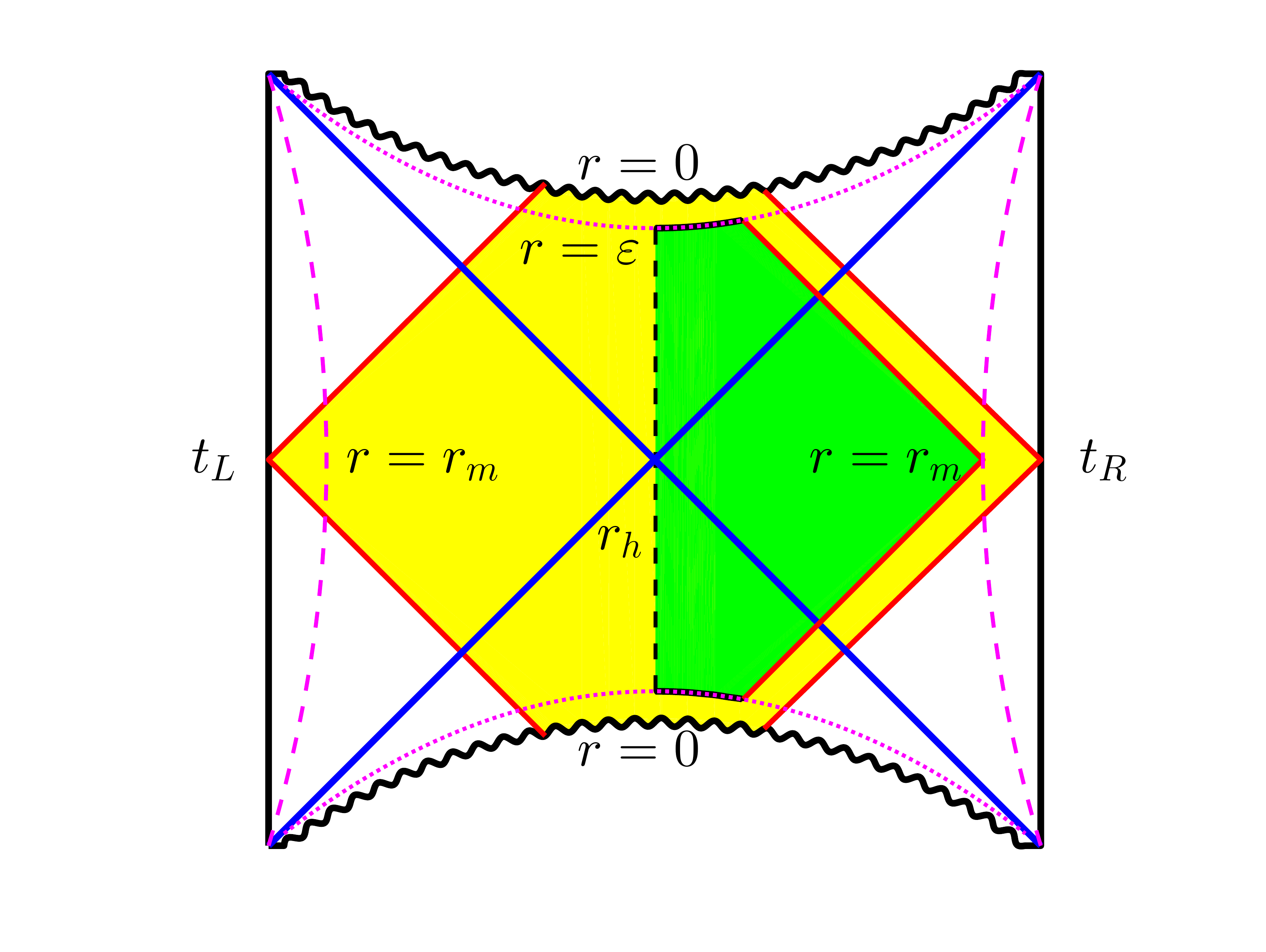}
  \caption{The penrose diagram and the regularized approach of the WDW patch for $t_R=t_L=0$ in the Schwarzschild AdS black holes. The null boundaries of the WDW patch come from the finite cut-off boundary and there is a null-null joint at the cut-off $r=r_m$. In addition, in order to regularize the singularity, we need to use an additional cut-off at $r=\varepsilon\rightarrow0$, so there are also some new joints and space-like boundaries.  }\label{FigSAdS1}
\end{figure}
Similar to the case in BTZ black hole, there is a null-null joint at the cut-off $r=r_m$.  When $r_m=\infty$, the null sheets coming from the boundaries will meet the singularity before they meet each others. In order to regularize the singularity, we need to use an additional cut-off at $r=\varepsilon\rightarrow0$, so there are also some new joints and space-like boundaries. (See Fig.~\ref{FigSAdS1}.)

The null boundaries at the green region in Fig.~\ref{FigSAdS1} is given by $v=v_m$ and $u=u_m$, where $v_m \equiv v(0,r_m)=r^*(r_m)$ and $u_m\equiv u(0,r_m)=-r^*(r_m)$.  The dual normal vectors for such null boundaries are still given by $k_I=\alpha[(\td t)_I+r^{-2}f^{-1}(\td r)_I]$ and $\bar{k}_I=\beta[(\td t)_I-r^{-2}f^{-1}(\td r)_I]$. Here we still explicitly exhibit the freedom of choosing the dual normal vector by two arbitrary constant $\alpha$ and $\beta$.

The bulk action is expressed as
\begin{equation}\label{AdS4bulk}
\begin{split}
  I_{\text{bulk}}=&-\frac{3\Sigma_{2,k}}{2\pi\el^2}\int_0^{r_m}[v_m-r^*(r)]r^2\td r\\
  =&-\frac{\Sigma_{2,k}}{4\pi}r_m^2+\frac{k\Sigma_{2,k}\el^2}{2\pi}\ln(r_m/\el)+I_0+\mathcal{O}(\el/r_m)\,.
  \end{split}
\end{equation}
Here $I_0$ is the finite term, which reads
\begin{equation}\label{SAdSfinitI0}
\begin{split}
  I_0&=-\frac{\Sigma_{2,k}d}{2\pi}\left\{\frac{k^2\el^4+3k\el^2r_h^2+r_h^4}{6(k\el^2+3r_h^2)}\ln\left(\frac{k^2\el^2+r_h^2}{\el^2}\right) -\frac{r_h^4}{3(k\el^2+3r_h^2)}\ln\left(\frac{r_h}{\el}\right)\right.\\
  &+\left.\frac{r_h(k^2\el^4+5k\el^2r_h^2+3r_h^4)}{3(k\el^2+3r_h^2)\sqrt{4k\el^2+3r_h^2}}\left[\frac{\pi}2-\arctan\left(\frac{r_h}{\sqrt{4k\el^2+3r_h^2}}\right)\right] \right\}.
   \end{split}
\end{equation}
We see that a logarithm term appears in Eq.~\eqref{AdS4bulk}. As  the null-spacelike joints at the corners $r=0$ have no contributions on the action \cite{Chapman:2016hwi}. The joint term at the infinite boundary for general $d$ is given by following expression
\begin{equation}\label{SAdSjoint}
\begin{split}
  I_{\text{joint}}&=\left.\frac{\Sigma_{d-1,k}}{4\pi}r^{d-1}\ln(|k^\mu\bar{k}_\mu|/2)\right|_{r=r_m}=\frac{\Sigma_{d-1,k}}{4\pi}r_m^{d-1}\ln\left[\frac{r_m^2f(r_m)}{\alpha\beta}\right]\\
  &=\frac{\Sigma_{d-1,k}}{4\pi}r_m^{d-1}\ln\left[\frac{r_m^2}{\el^2\alpha\beta}\right]+\frac{\Sigma_{d-1,k}}{4\pi}r_m^{d-1}\ln\left(1+\frac{k\el^2}{r_m^2}-\frac{\omega_{d-2}}{r^d}\right)\\
  &=\frac{\Sigma_{d-1,k}}{4\pi}r_m^{d-1}\ln\left[\frac{r_m^2}{\el^2\alpha\beta}\right]+\frac{\Sigma_{d-1,k}}{4\pi}r_m^{d-1}\left(\frac{k\el^2}{r_m^2}-\frac{k^2\el^4}{2r_m^4}+\cdots\right)\,.
  \end{split}
\end{equation}
One can check that $k_{I}$ is still affine parameterized, so we still find that only the null boundary term shown in Eq. \eqref{addnulbd} has contribution. The expansions of $k_\mu$ and $\bar{k}_I$ in general $d$ are
\begin{equation}\label{expnsionSAdS}
  \Theta=g^{IJ}\nabla_I k_{J}=\frac{(d-1)\alpha}r,~~\bar{\Theta}=g^{IJ}\nabla_I \bar{k}_{J}=-\frac{(d-1)\beta}r.
\end{equation}
By the similar method in Eq.~\eqref{IlambdaBTZ}, we find the null boundary term $I_\lambda$ is
\begin{equation}\label{IlambdaAdS}
  I_\lambda=\frac{\Sigma_{d-1,k}}{4\pi}\left\{[2\ln(d-1)+\ln(\alpha\beta\el^2/r_m^2)+\frac2{d-1}]r_m^{d-1}\right\}.
\end{equation}

An important difference between the Schwarzschild black hole and the BTZ black hole is that there is a space-like curvature singularity at $r=0$. We need to make a cut-off at $r=0$ so that the computation cannot touch the singularity. As a result, there are two space-like surface terms at $r=\varepsilon\rightarrow0$. The contribution of such terms on the action can be given the similar method shown Ref.~\cite{Chapman:2016hwi}\footnote{However, there is a little difference between our result and the result in Ref.~\cite{Chapman:2016hwi}. In Ref.~\cite{Chapman:2016hwi}, the null sheets come from the boundary $r=\infty$. Here the null sheets come from  the cut-off surface $r=r_m$}. For the case, $d=3$, it is
\begin{equation}\label{GHKSAdS}
\begin{split}
  I_{\text{GHY}}&=\frac{3\Sigma_{2,k}}{4\pi}\omega_1(v_m-r^*(0))=\frac{3\Sigma_{2,k}}{4\pi}\left\{\frac{r_h^2(k\el^2+r_h^2)}{2(k\el^2+3r_h^2)}\ln\left(\frac{k^2\el^2+r_h^2}{r_h^2}\right)\right.\\
   &+\left.\frac{(k\el^2+r_h^2)(2k\el^2+3r_h^2)r_h}{(k\el^2+3r_h^2)\sqrt{4k\el^2+3r_h^2}}\arctan\sqrt{\frac{4k\el^2}{r_h^2}+3}\right\}+\mathcal{O}(\el/r_m).
  \end{split}
\end{equation}

For the case that $d>2$, the surface counterterm is nonzero.  We see $F_{A}^{0}=\ln(d-1)/(4\pi)$. It is easy to see that $K_{ij}=0$ and $R=\hat{R}=k(d-1)(d-2)/r^2$.
Specializing that $d=3$, there is a logarithm counterterm in the subleading counterterm.
\begin{equation}\label{SAdSlog}
  \frac{k\Sigma_{2,k}\el^2}{4\pi}\ln(r_m/\el)\,.
\end{equation}
Thus, the surface counterterm for $d=3$ reads
\begin{equation}\label{surfSAdSA}
  I_{\text{ct,L}}=I_{\text{ct,R}}=\frac{\Sigma_{2,k}\ln2}{4\pi}r_m^2+\frac{k\Sigma_{2,k}\el^2}{4\pi}\ln(r_m/\el)\,.
\end{equation}
Finally, we obtain the regularized complexity for $d=3$
\begin{equation}\label{SAdSCA}
\begin{split}
  \mathcal{C}_{A,\text{reg}}&=\frac1{\pi\hbar}\lim_{r_m\rightarrow\infty}(I_{\text{bulk}}+I_{\text{GHY}}+I_{\text{joint}}+I_{\lambda}-I_{\text{ct,L}}-I_{\text{ct,R}})\\
  &=\frac{\Sigma_{2,k}}{4\pi^2\hbar}\left\{\frac{r_h^4-2k^2\el^4-3k\el^2r_h^2}{2(k\el^2+3r_h^2)}\ln\left(k+\frac{r_h^2}{\el^2}\right)-\frac{r_h^2(r_h^2+3k\el^2)}{(k\el^2+3r_h^2)}\ln\left(\frac{r_h}{\el}\right)\right.\\ &\left.+\frac{r_h(4k^2\el^4+5k\el^2r_h^2+3r_h^4)}{(k\el^2+3r_h^2)\sqrt{4k\el^2+3r_h^2}}\arctan\sqrt{\frac{4k\el^2}{r_h^2}+3}\right\}\\
  &+\frac{\Sigma_{2,k}k\el^2}{4\pi^2\hbar}\,.
  \end{split}
\end{equation}
As we expected, all the divergent terms have disappeared and the result is independent of the values of $\alpha$ and $\beta$ when we choose the null normal vectors for the null boundaries.

Though Eq.~\eqref{SAdSCA} is obtained by the assumption $r_h>2\el/\sqrt{3}$ when $k=-1$, we make an analytical extension to get the regularized complexity when $\el<r_h<2\el/\sqrt{3}$  by following analytical extension
\begin{equation}\label{analytical}
  \sqrt{3r_h^2-4\el^2}=i\sqrt{4\el^2-3r_h^2},~~\arctan\sqrt{3-\frac{4\el^2}{r_h^2}}=i\text{arctanh}\sqrt{\frac{4\el^2}{r_h^2}-3}\,.
\end{equation}
On the other hand, by the following identity for arctanh function and logarithm function when $x>1$
\begin{equation}\label{arctanhlog}
  \text{arctanh}(x)=\frac12[\ln(1+x)-\ln(1-x)]\,,
\end{equation}
one can check that the Eq.~\eqref{SAdSCA} has well defined limit at $r_h=\el$ and is analytical in the neighbourhood of $r_h=\el+0^+$. So the Eq.~\eqref{SAdSCA} can extend into the whole region of $r_h\geq\el$ when $k=-1$. By this analytical extension, it is easy to find that the vacuum regularized complexity for $k=0,1$($r_h=0$) and $k=-1$($r_h=\el$) is
\begin{equation}\label{SAdSvaccumA}
   \mathcal{C}_{\text{A,reg,vac}}=\frac{\Sigma_{2,k}k\el^2}{4\pi^2\hbar}\,.
\end{equation}

\kyr{We plot the regularized complexity \eqref{SAdSCA} and \eqref{SAdSvaccumA} in Fig. \ref{Fig:RegC}.  They
may not be positive but their difference, the complexity of formation, is always positive and
the same as the results in Ref.~\cite{Chapman:2016hwi}.} We note that the complexity of formation for $k=-1$ and $\el<r_h<2\el/\sqrt{3}$ has also been given by Ref.~\cite{Chapman:2016hwi} in a very implicit manner. In fact, one can prove that it is just  the same as the Eq.~\eqref{SAdSCA} in the sense of analytical extension shown in \eqref{analytical}.

When $\el/\sqrt{3}<r_h<\el$ and $k=-1$, i.e., the small black hole case in hyperbolic black holes, the logarithm function and arctanh function become multiple values and, the casual structure of such hyperbolic black hole is very different from what we have shown in the Fig.~\ref{FigSAdS1}. In principle, we need an additional computation for this case.  We leave this case in future works.

\subsubsection*{Case of $d=4$}
When $d=4$, we see that the logarithm term will not appear but the subleading counterterm appears. By Eq.~\eqref{CAdiv111} and \eqref{expFA2}, the total surface counterterm reads
\begin{equation}\label{subF2SAdS}
  I_{\text{ct,L}}= I_{\text{ct,R}}=\frac{\Sigma_{3,k}}{4\pi}(r_m^3\ln3+\frac32kr_m\el^2)\,.
\end{equation}
The bulk   can be computed by the same method shown in Eq.~\eqref{BTZbulk} and  the result is
\begin{equation}\label{AdS5bulk}
  I_{\text{bulk}}=-\frac{4\Sigma_{3,k}}{2\pi\el^2}\int_0^{r_m}[v_m-r^*(r)]r^2\td r=-\frac{\Sigma_{3,k}}{6\pi}r_m^3+\frac{k\Sigma_{3,k}\el^2}{2\pi}r_m+I_0+\mathcal{O}(\el/r_m),
\end{equation}
where
\begin{equation}\label{AdS4finite}
  I_0=-\frac{\Sigma_{3,k}}{4}\frac{(k\el^2+r_h^2)^{5/2}}{k\el^2+2r_h^2}\,.
\end{equation}
And the contribution of boundary terms coming from the singularity is
\begin{equation}\label{GHKSAdS4}
  I_{\text{GHY}}=\frac{\Sigma_{3,k}}2\frac{r_h^2(r_h^2+k\el^2)^{3/2}}{2r^2+k\el^2}.
\end{equation}

\begin{figure}[]
	\begin{center}
		      \subfigure[CA conjecture: Solid lines are \eqref{SAdSCA} and \eqref{SAdS5CA}.    Dashed lines are \eqref{SAdSvaccumA} and \eqref{AdS5vca}. ]
			{
                           \includegraphics[width=.4\textwidth]{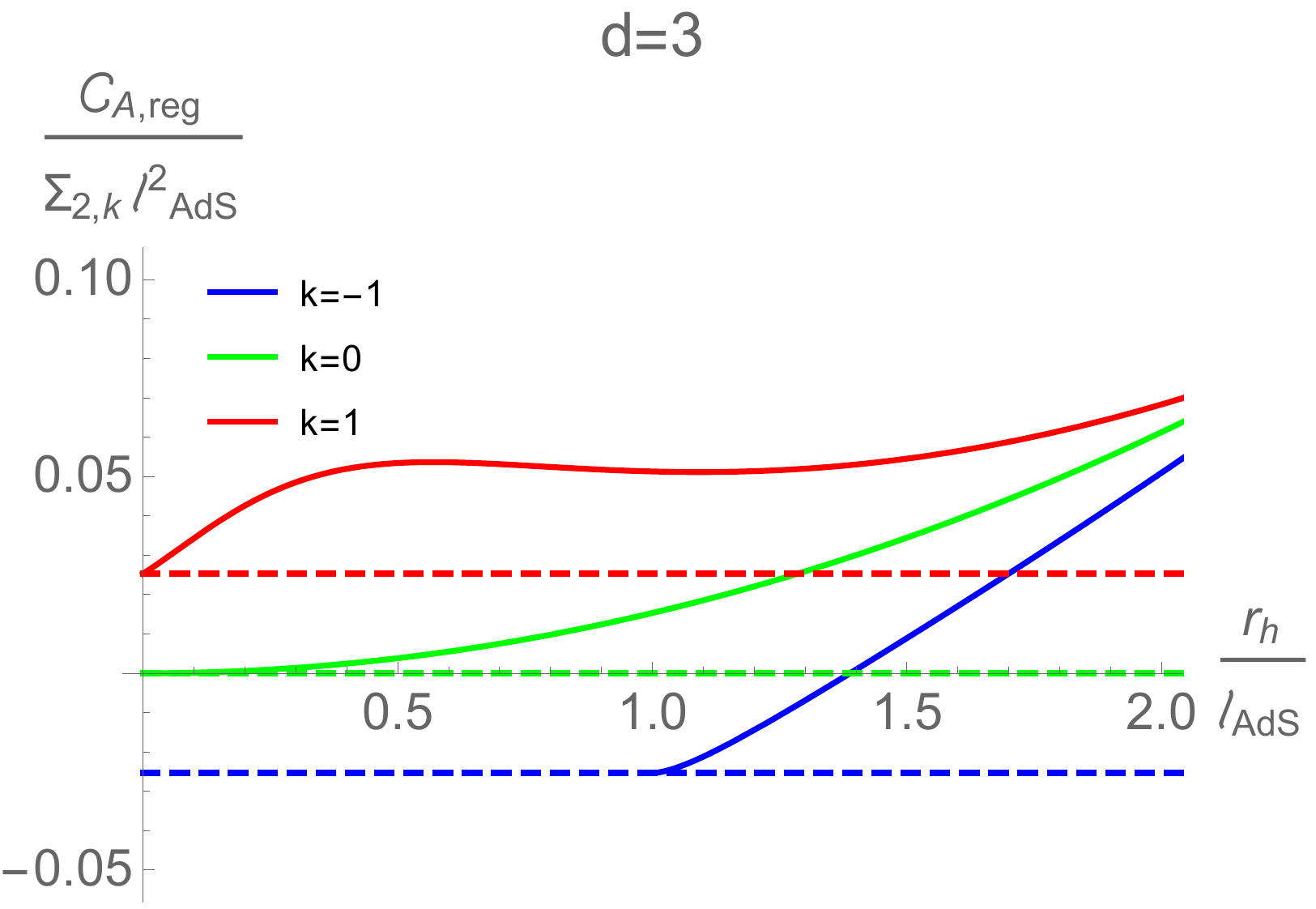} \ \ \ \ \
                           \includegraphics[width=.4\textwidth]{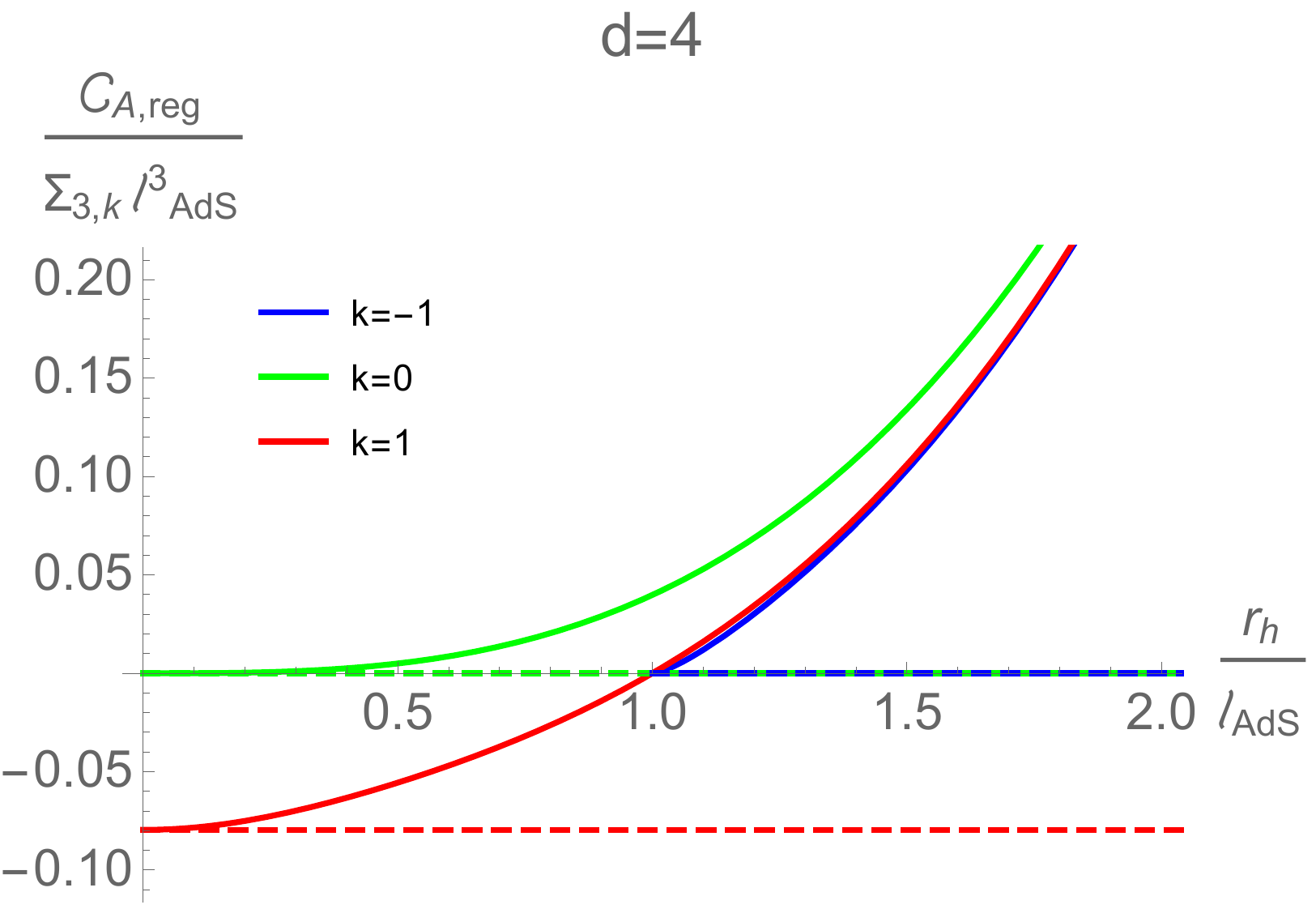}
			}
	              \subfigure[CV conjecture: Solid lines are \eqref{po1} and \eqref{po3}.    Dashed lines are \eqref{po2} and \eqref{po4}.  For $k=0$ see \eqref{po5}. ]
			{
                              \includegraphics[width=.4\textwidth]{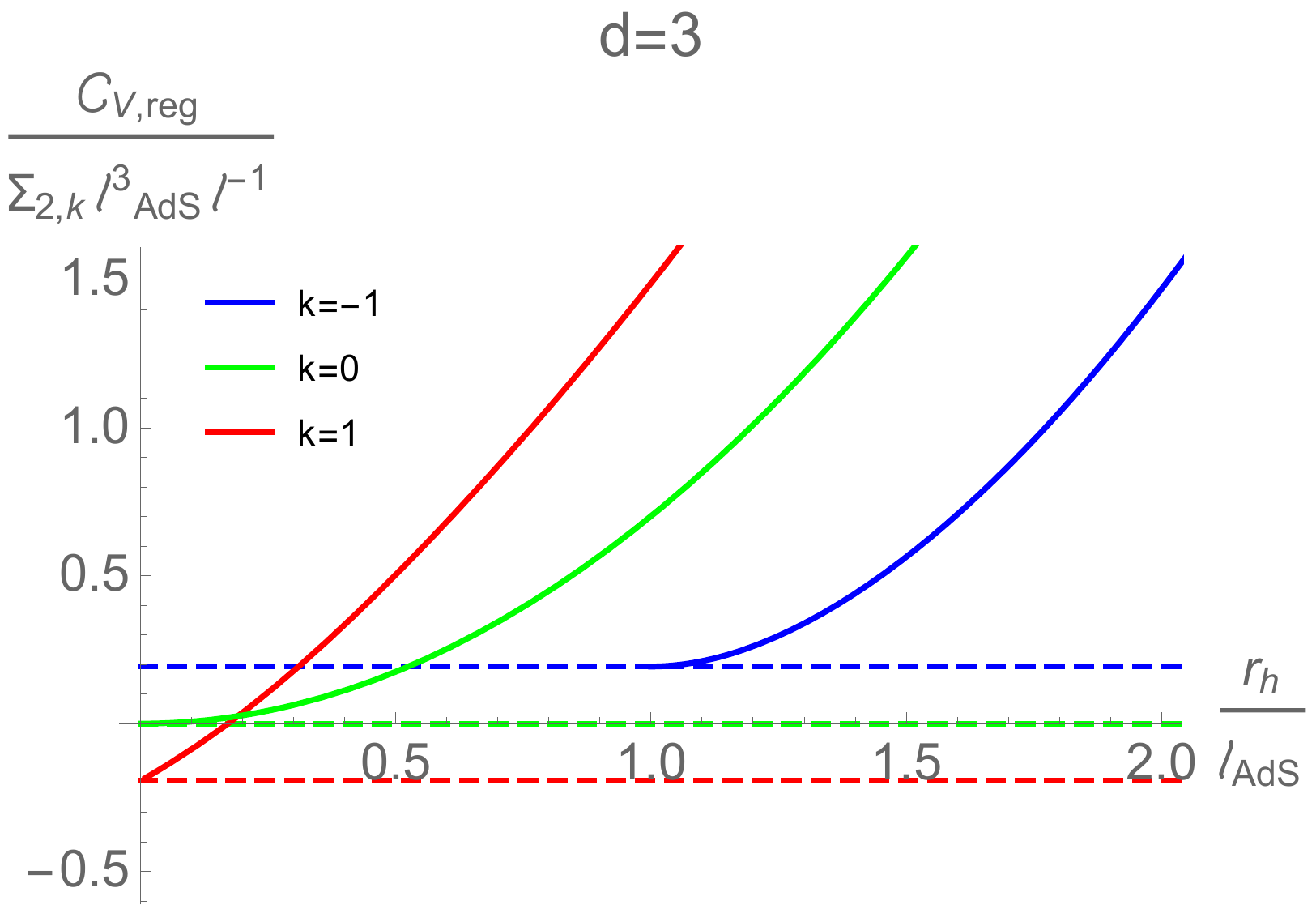} \ \ \ \ \
                             \includegraphics[width=.4\textwidth]{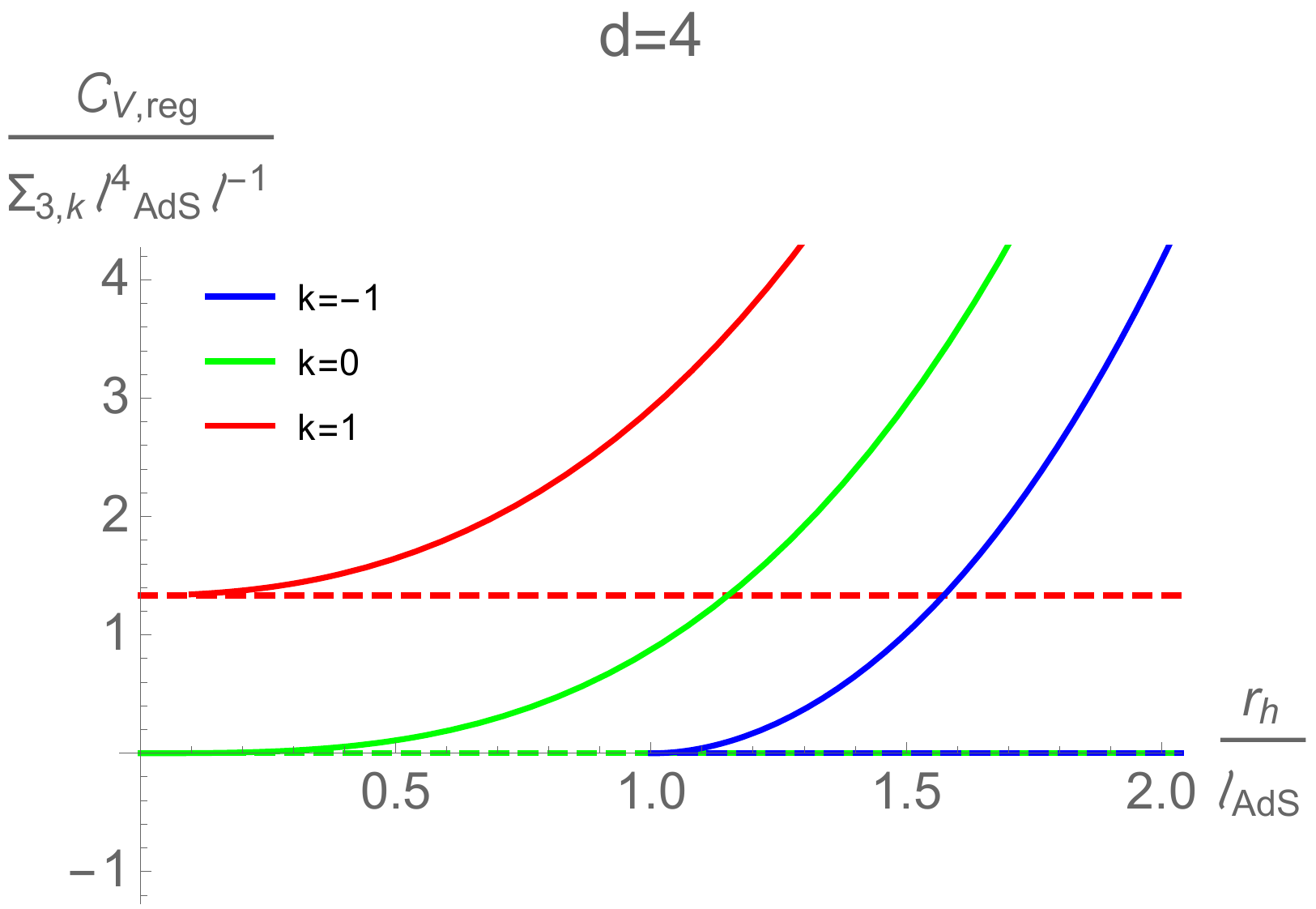}
                             }
	\end{center}
  \caption{Regularized complexity. The regularized complexity may be negative but the complexity of formation (solid line minus dashed line) is always positive, which agree to Fig. 4,5,10 and 11 in \cite{Chapman:2016hwi}. }\label{Fig:RegC}
\end{figure}

The joint terms and null boundary terms $I_\lambda$ can be obtained by Eq.~\eqref{SAdSjoint} and \eqref{IlambdaAdS} with $d=4$. Then we find that all the divergent terms can be canceled with each other and we obtain a finite regularized complexity
\begin{equation}\label{SAdS5CA}
  \mathcal{C}_{\text{A,reg}}=\frac{\Sigma_{3,k}}{4\pi\hbar}\frac{(k\el^2+r_h^2)^{3/2}(r_h^2-k\el^2)}{k\el^2+2r_h^2}\,.
\end{equation}
By this result, we can obtain the vacuum regularized complexity
\begin{equation}\label{AdS5vca}
  \mathcal{C}_{\text{A,reg,vac}}=-\frac{\Sigma_{3,k}}{4\pi\hbar}\el^3\delta_{k,1}.
\end{equation}
\kyr{We plot the regularized complexity \eqref{SAdS5CA} and \eqref{AdS5vca} in Fig. \ref{Fig:RegC}.  They
may not be positive but their difference, the complexity of formation, is always positive and
the same as the results in Ref.~\cite{Chapman:2016hwi}.}
 Similarly, the Eq.~\eqref{SAdS5CA} is valid when $r_h\geq\el$ in hyperbolic black holes. The case of small black hole needs another computation.

\subsection{Regularized complexity in CV conjecture}
Now let us calculate the regularized complexity for the CV conjecture. For the case $t_L=t_R=0$, the maximal valume surface bounded by codimension-two surface $t_L$ and $t_R$ is just the time slice of  $t=0$. Then volume of this codimension-one surface can be obtained by the following integration
\begin{equation}\label{}
\begin{split}
V&=2\Sigma_{d-1,k}\int_{r_h}^{\infty}\frac{r^{d-2}}{\sqrt{f(r)}}\td r\\
&=2\Sigma_{d-1,k}\int_{r_h}^{\infty}r^{d-2}\left(\frac{k}{r^2}+\frac{1}{\el^2}-\frac{r_h^{d}}{r^d}\left(\frac{k}{r_h^2}+\frac1{\el^2}\right)\right)^{-1/2}\td r  \,,
\end{split}
\end{equation}
of which near boundary ($r \rightarrow \infty $) behaviour is
\begin{equation} \label{kkkkk1}
2\Sigma_{d-1,k}\el\int_{r_h}^{\infty}\left(r^{d-2}-\frac{k\el^2}{2}r^{d-4}+\cdots\right)\td r.
\end{equation}
We will give the regularized complexity in different dimension and $k$.
\subsubsection*{Planar Geometry}
In this case, $k=0$ and the boundary is just a flat space-time, which leads that $R_{\mu\nu}=K_{ij}=0$ at the boundary. We can obtain the results for general dimension.
Because the divergence structure \eqref{kkkkk1} is
\begin{equation}
2\Sigma_{d-1,0}\el\int_{r_h}^{r_m}\left(r^{d-2}+\frac{r_h^d}{2r^2}+\cdots\right)\td r \,,
\end{equation}
the volume of the maximal surface at the cut-off $r_m=\el/\delta$  is
\begin{equation}\label{}
\begin{split}
V_\delta &=2\Sigma_{d-1,0}\el\left(\frac{\el^{d-1}}{(d-1)\delta^{d-1}}+\int_{r_h}^{r_m}\left(\frac{r^{d-1}}{\sqrt{r^2-\frac{r_h^d}{r^{d-2}}}}-r^{d-2}\right)\td r-\frac{r_h^{d-1}}{d-1}\right)\\
&=2\Sigma_{d-1,0}\el\left(\frac{\el^{d-1}}{(d-1)\delta^{d-1}}+\frac{\sqrt{\pi}(d-2)\Gamma(1+\frac{1}{d})}{2(d-1)\Gamma(\frac1{2}+\frac{1}{d})}r_h^{d-1}\right).
\end{split}
\end{equation}
The surface counterterms are
\begin{equation}\label{CVplanar}
\begin{split}
V^{(1)}_{\text{ct}}&=\frac{\el}{d-1}\int_B\td^{d-1} x\sqrt{\sigma}=\frac{\Sigma_{d-1,k}\el^d}{(d-1)\delta^{d-1}}\,,\\
V^{(n)}_{\text{ct}}&=0,~~~~n>1.
\end{split}
\end{equation}
Then the regularized complexity can be written as
\begin{equation}\label{po5}
\begin{split}
\mathcal{C}_{\text{V,reg}}&=\frac{1}{\ell}\lim_{\delta\rightarrow0}(V-2V^{(1)}_{\text{ct}})\\
&= \frac{\el}{\ell}\frac{\Sigma_{d-1,0}\sqrt{\pi}(d-2)\Gamma(1+\frac{1}{d})}{(d-1)\Gamma(\frac1{2}+\frac{1}{d})}r_h^{d-1}\,,
\end{split}
\end{equation}
\kyr{and $\mathcal{C}_{\text{V,reg,vac}} = 0$.  We plot the regularized complexity \eqref{po5} for $d=3,4$ in Fig. \ref{Fig:RegC}.  The complexity of formation is always positive and  the same as the results in Ref.~\cite{Chapman:2016hwi}.}

\subsubsection*{Spherical and Hyperbolic Geometries for $d=3$}

In this case, the divergence structure \eqref{kkkkk1} is
\begin{equation}
2\Sigma_{2,k}\el\int_{r_h}^{r_m}\left(r-\frac{k\el^2}{2r}+\cdots\right)\td r \,,
\end{equation}
so we have
\begin{equation}\label{}
\begin{split}
V_\delta=&2\Sigma_{2,k}\el\left[\frac{\el^2}{2\delta^2}+\frac{k\el^2}{2}\text{ln}\delta-\frac{r_h^2}{2}+\frac{k\el^2}{2}\ln(r_h/\el)\right.\\
&+\left.\int_{r_h}^{r_m}\left(\frac{r^2}{\sqrt{k\el^2+r^2-\frac{r_h}{r}(k\el^2+r_h^2)}}-r+\frac{k\el^2}{2r}\right)\td r\right]\,,
\end{split}
\end{equation}
where  $r_m=\el/\delta$.
Now we need the first order surface counterterm and the subleading logarithmic counterterm:
\begin{equation}\label{CVknonzero}
\begin{split}
V^{(1)}_{\text{ct}}&=\frac{\el}{2}\int_B\td^2 x\sqrt{\sigma}=\frac{\Sigma_{2,k}\el^3}{2\delta^2} \,, \\
V_{\text{ct}}^{\text{(2)}}&=\frac{k\Sigma_{2,k}\el^3}{2}\text{ln}\delta\,.
\end{split}
\end{equation}
Thus, the regularized complexity can be written as
\begin{equation}\label{po1}
\begin{split}
\mathcal{C}_{\text{V,reg}}&=\lim_{\delta\rightarrow0}\frac1{\ell}(V_\delta-2V^{(1)}_{\text{ct}}-2V_{\text{ct}}^{\text{(2)}})\\
&=2\Sigma_{2,k}\el^3\ell^{-1}\left(\int_{x_h}^{\infty}\left(\frac{x^2}{\sqrt{k+x^2-\frac{x_h}{x}(k+x_h^2)}}-x+\frac{k}{2x}\right)\td x-\frac{x_h^2}{2}+\frac{k}{2}\text{ln}x_h\right).
\end{split}
\end{equation}
Here we define $x=r/\el$ and $x_h=r_h/\el$. All the divergent terms have disappeared, as expected. It is straightforward to find that the vacuum regularized complexity is
\begin{equation}\label{po2}
\mathcal{C}_{\text{V,reg,vac}}=-k\ell^{-1}\Sigma_{2,k}\el^3\left(\text{ln}2-\frac{1}{2}\right).
\end{equation}
\kyr{We plot the regularized complexity \eqref{po1} and \eqref{po2} in Fig. \ref{Fig:RegC}.  They
may not be positive but their difference, the complexity of formation, is always positive and
the same as the results in Ref.~\cite{Chapman:2016hwi}.}

\subsubsection*{Spherical and Hyperbolic Geometries for $d=4$}
In this case, the divergence structure \eqref{kkkkk1} is
\begin{equation}
2\Sigma_{3,k}\el\int_{r_h}^{r_m}\left(r^2-\frac{k\el^2}{2}+\cdots\right)\td r \,,
\end{equation}
so we have
\begin{equation}\label{}
\begin{split}
V_\delta &=2\Sigma_{3,k}\el^4\left[\frac{1}{3\delta^3}-\frac{k}{2\delta}+\int_{x_h}^{\infty}\left(\frac{x^3}{\sqrt{k+x^2-\frac{x_h^2}{x^2}(k+x_h^2)}}-x^2+\frac{k}{2}\right)\td x-\frac{x_h^3}{3}+\frac{kx_h}{2}\right]\,,
\end{split}
\end{equation}
where  $r_m=\el/\delta$, $x=r/\el$ and $x_h=r_h/\el$. In this case we need the first and second surface counterterms, which are
\begin{equation}\label{CVknonzero2}
\begin{split}
V^{(1)}_{\text{ct}}&=\frac{\el}{3}\int_B\td^3 x\sqrt{\sigma}=\frac{\Sigma_{3,k}\el^4}{3\delta^3} \,, \\
V^{(2)}_{\text{ct}}&=-\frac{\el^3}{4}\int_B\td^3 x\sqrt{\sigma}\left[\frac{2}{3}(\hat{R}-R/2)-\dfrac{4}{9}K^2\right]=-\frac{k\Sigma_{3,k}\el^4}{2\delta}\,.
\end{split}
\end{equation}
Then the regularized complexity can be written as
\begin{equation}\label{po3}
\begin{split}
\mathcal{C}_{\text{V,reg}}&=\lim_{\delta\rightarrow0}\frac1{\ell}(V_\delta-2V^{(1)}_{\text{ct}}-2V^{(2)}_{\text{ct}})\\
&=2\Sigma_{3,k}\el^4\ell^{-1}\left(\int_{x_h}^{\infty}\left(\frac{x^3}{\sqrt{k+x^2-\frac{x_h^2}{x^2}(k+x_h^2)}}-x^2+\frac{k}{2}\right)\td x-\frac{x_h^3}{3}+\frac{kx_h}{2}\right).
\end{split}
\end{equation}
All the divergent terms have disappeared and the vacuum regularized complexity is
\begin{equation}\label{po4}
\mathcal{C}_{\text{V,reg,vac}}=\left\{
\begin{split}
&\frac{4}{3\ell}\Sigma_{3,k}\el^4,~~~~k=1,\\
&0,~~k=-1.
\end{split}
\right.
\end{equation}
%

\kyr{We plot the regularized complexity \eqref{po3} and \eqref{po4} in Fig. \ref{Fig:RegC}.  They
may not be positive but their difference, the complexity of formation, is always positive and
the same as the results in Ref.~\cite{Chapman:2016hwi}.}



\section{Summary}\label{Summary}
In this paper, we studied how to obtain the finite term in a covariant manner from the holographic complexity  for both CV and CA conjectures {when the boundary geometry is not deformed by relevant operators}. Inspired by the recent results that the divergent terms are  determined only by the boundary metric and have no relationship to the stress tensor and bulk matter fields, we showed that such divergences can be canceled by adding codimension-two boundary counterterms.  If bulk dimension is even, a logarithmic divergence appears.  These boundary surface counterterms do not contain any boundary stress tensor information so they are non-dynamic background and can be subtracted from the complexity without any physical effects.
In the CA conjecture, with the modified boundary term proposed by Ref.~\cite{Reynolds:2016rvl} different from the framework  in the Ref.~\cite{Chapman:2016hwi,Carmi:2016wjl}, our regularized complexity is also independent on the choice of the normalization of the affine parameters of the null normal vectors.
We argue that the regularized complexity for both CV and CA conjectures contain all the information of dynamics and matter fields in the bulk for given time slices, and we can use them to study the dynamic properties of the holographic complexity such as the growth rate and the complexity of formation.

We  showed the  minimal subtraction counterterms for both CA and CV conjectures up to the dimension $d+1\leq5$. By these surface conunterterms, we calculated the regularized complexity for the non-rotational and rotational BTZ black holes and  the Schwarzschild AdS black holes in four and five dimensions with different horizon topologies. They also directly show that the problem that the complexity depends on the choice of the normalization about the null normal vectors in the CA conjecture will not appear in the regularized complexity.
As a check, we use our regularized complexity to compute the complexity of formation in the BTZ black holes and  the AdS$_{d+1}$ black holes  based on both CA and CV conjectures and reproduced  the same results shown in Ref.~\cite{Chapman:2016hwi}. However, unlike Ref. \cite{Chapman:2016hwi} we do not need to worry about the coordinate dependence of the cut-off, because our regularized complexity is defined to be coordinate independent.

Using this regularized complexity, we can study the effects of bulk matter fields and thermodynamic conditions on the holographic complexity at a fixed dual boundary (the codimension-one surface) geometry.   There are many future works. For example, we can study its behavior in holographic superconductor models to see if it can play a role of an order parameter in phase transitions or if there is any interesting and special behavior at  zero temperature limit~\cite{Yang1}. We also can directly compute the complexity at different time slices and compute its derivative with respective to $t_L$ or $t_R$  rather than only the case $t_L=t_R$ shown in the examples in this paper and obtain the whole growth rate if $(\partial/\partial t)^\mu$ is a timelike Killing vector at the boundaries.

\kyr{In this paper, it is assumed that the asymptotic boundary geometry has an expansion shown in Eq. \eqref{powerz1}.
However, it can be deformed by a relevant operator, for example by a scalar field with negative mass.\footnote{We thank the anonymous referee to draw our attention to this issue.} 
In such a deformed metric, the divergent structure will depend also on the information of the matter field, so our formalism cannot cancel all the divergences in both CV and CA conjectures. It would be interesting to analyse the UV divergent structures in this case and find the counterterms. We are now investigating this problem. }

\acknowledgments

We would like to thank Rob Myers for valuable discussions and correspondence. We also thank Yong-Jun Ahn for plotting Fig. 6. The work of K.Y.Kim and C. Niu  was supported by
Basic Science Research Program through the National Research Foundation of Korea(NRF) funded by the Ministry of Science, ICT $\&$ Future Planning(NRF- 2017R1A2B4004810) and GIST Research Institute(GRI) grant funded by the GIST in 2017.
We also would like to thank ``11th Asian Winter School on Strings, Particles, and Cosmology'' at Sun Yat-Sen university in Zuhai, China for the hospitality during our visit, where part of this work was done.

\appendix
\section{Subleading divergent terms in CA conjecture}\label{app1}
In this appendix, we will show how to obtain the contribution of the null surface term coming from the nonzero $\kappa$ in the action~\eqref{actionull} in more detail. {It seems that Refs.~\cite{Carmi:2016wjl, Reynolds:2016rvl} neglected an $\mathcal{O}(z^3)$ order contribution from the null boundary contribution, so our result is slightly different from theirs.}\footnote{{Recently, we learned that the authors of \cite{Carmi:2016wjl} obtained
the same results as ours by a different method and it would be updated in their revised version. We thank the authors of \cite{Carmi:2016wjl} for sharing the manuscript before posting.}}

Based on Ref.~\cite{Lehner:2016vdi}, the null boundary term in the action can be written as
\begin{equation}\label{nullbdterm}
  I_\mathcal{N}=-\frac1{8\pi}\int_\mathcal{N}\sqrt{\sigma}\kappa\td\lambda\td^2x\,.
\end{equation}
Here $k^I=(\partial/\partial\lambda)^I$ is the normal vector of the null surface, $\lambda$ is the parameter of integral curve of $k^I$. Following the Ref.~\cite{Carmi:2016wjl}, we assume $k_\mu$ has  the following form near the boundary
\begin{equation}\label{formkmu}
  k_I=\alpha(-dz+n_\mu\td x^\mu)\,.
\end{equation}
Here $\alpha$ is a constant but $n_\mu$ is the function of $z$ and $x^\mu$. Using the metric ~\eqref{FGmetric}, we find that
\begin{equation}\label{formkmu2}
  k^I=\frac{\alpha z^2}{\el^2}(-\partial_z+n^\mu\partial_\mu)\,,
\end{equation}
where $n^\mu=g^{\mu\nu}n_\nu$. Because the Eq.~\eqref{formkmu} is not an additional assumption we can always write the normal vector for the null surface as Eq.~\eqref{formkmu2} in the FG coordinate system. The null condition $k_I k^I=0$ shows that $n_\mu$ must be a normalized unit time-like vector, i.e.
\begin{equation}\label{eqnmu1}
  n_I n^I=-1\,.
\end{equation}

Now let us find the {non-affinity parameter} $\kappa$ for this null normal vector. Using $k^\mu\nabla_\mu k^\nu=\kappa k^\nu$ we have
\begin{equation}\label{geodisceq1}
  \frac{\td k^I}{\td\lambda}+{\Gamma^I}_{JK}k^J k^K=\kappa k^I \, .
\end{equation}
We will solve this equation order by order in $z$. One can see that under the gauge $\tilde{N}^{(0)} = 1$ and $\tilde{L}^{i(0)} = 0$ in Eq. \eqref{boundmetric}, $n^\mu$ must have  the following form
\begin{equation}\label{expandnmu1}
  n^\mu=-\delta^\mu_t+n^{\mu(1)}z^2+n^{\mu(2)}z^4+\cdots \,,
\end{equation}
and $\kappa$ has the following series expansion with respective to $z$
\begin{equation}\label{expandnmu1}
  \kappa=\kappa^{(0)}+\kappa^{(1)}z^3+\cdots\,.
\end{equation}
Here the coefficients $\{n^{\mu(1)},n^{\mu(2)},\cdots\}$ and $\{\kappa^{(0)}, \kappa^{(1)}, \cdots\}$ are  only the function of $x^\mu$.

As the Eq.~\eqref{formkmu} shows that  $\td z/\td\lambda=k^z$ and $\td x^\mu/\td\lambda=k^\mu$ on the integral curve of $k^I$, we can use the following replacement when we compute the integral \eqref{nullbdterm}
\begin{equation}\label{dlambda2dz}
  \frac{\td}{\td\lambda}=\frac{\alpha z^2}{\el^2}\left(-\frac{\partial}{\partial z}+n^\mu\frac{\partial}{\partial x^\mu}\right),\qquad \td\lambda=-\frac{\el^2}{\alpha z^2}\td z\,.
\end{equation}
Therefore,
\begin{equation}\label{dkdlambdazt}
  \begin{split}
   \frac{\td k^z}{\td\lambda}&=\frac{2\alpha^2z^3}{\el^4},\\
   \frac{\td k^t}{\td\lambda}&=\frac{2\alpha^2z^3}{\el^4}-\frac{4\alpha^2n^{t(1)}}{\el^4}z^5+\mathcal{O}(z^7),\\
   I_\mathcal{N}&=\frac{\el^2}{8\pi\alpha }\int_\mathcal{N}\sqrt{\sigma}\frac{\kappa}{z^2}\td z\td^2x\,.
  \end{split}
\end{equation}
The relevant components of connection  ${\Gamma^I}_{JL}$ in Eq.~\eqref{geodisceq1}   are
\begin{equation}\label{comptsGamma}
\begin{split}
  &{\Gamma^z}_{zz}=-\frac1z,~~~~{\Gamma^z}_{\mu z}={\Gamma^z}_{z\mu}=0,~~~~{\Gamma^z}_{tt}=-\frac1z+\mathcal{O}(z^3),\\
  &{\Gamma^t}_{tt}=\mathcal{O}(z^2),~~~{\Gamma^t}_{tz}=-\frac1z-z\tilde{g}^{(1)}_{tt}+\mathcal{O}(z^3),~~~~{\Gamma^t}_{zz}=0\,.
  \end{split}
\end{equation}
so Eq.~\eqref{geodisceq1} reduces to
\begin{equation}\label{geodisc2}
\begin{split}
  &z:\quad \frac{2\alpha^2n^{t(1)}}{\el^4}z^5=-\frac{\alpha z^2}{\el^2}[\kappa^{(0)}+z^3\kappa^{(1)}],\\
  &t:\quad -2(\tilde{g}_{tt}^{(1)}+n^{t(1)})\frac{\alpha^2}{\el^4}z^5=\frac{\alpha z^2}{\el^2}[\kappa^{(0)}(n^{t(1)}z^2-1)-z^3\kappa^{(1)}]\,.
  \end{split}
\end{equation}
up to the order of $z^5$.
These give
\begin{equation}\label{order2kappa}
  \kappa^{(0)}=0,\quad \kappa^{(1)}=\frac{\alpha \tilde{g}^{(1)}_{tt}}{\el^2},\quad n^{t(1)}=-\frac{\tilde{g}^{(1)}_{tt}}2.
\end{equation}

Taking all these into account we evaluate the Eq.~\eqref{nullbdterm} as
\begin{equation}\label{intnull1}
\begin{split}
   I_\mathcal{N}&=\frac{1}{8\pi}\int\td^{d-1}x\int_\epsilon\sqrt{\sigma}\tilde{g}^{(1)}_{tt}z\td z\\
   &=\frac{1}{8\pi}\int\td^{d-1}x\sqrt{\tilde{\sigma}}\tilde{g}^{(1)}_{tt}\int_\epsilon z^{2-d}\td z+\mathcal{O}(\epsilon^{d-5})\\
    &=\frac{\epsilon^{d-3}}{8\pi(d-3)}\int_{z=\epsilon}\td^{d-1}x\sqrt{\tilde{\sigma}^{(0)}}\tilde{g}^{(1)}_{tt}+\mathcal{O}(\epsilon^{d-5})\,.
   \end{split}
\end{equation}
Using the equation
\begin{equation}\label{gtteq}
  \tilde{g}^{(1)}_{tt}=-\frac1{d-2}\left[\tilde{\hat{R}}^{(0)}-\frac{2d-3}{2(d-1)}\tilde{R}^{(0)}\right],
\end{equation}
we find the null boundary term contributes to the subleading divergence
\begin{equation}\label{intnull2}
\begin{split}
   I_\mathcal{N}&=-\frac{\epsilon^{d-3}}{8\pi(d-3)(d-2)}\int_B\td^2x\sqrt{\tilde{\sigma}^{(0)}}\left[\tilde{\hat{R}}^{(0)}-\frac{2d-3}{2(d-1)}\tilde{R}^{(0)}\right]\\
   &=-\frac{1}{8\pi(d-3)(d-2)}\int_B\td^2x\sqrt{\sigma}\left[\hat{R}-\frac{2d-3}{2(d-1)}R\right]+\mathcal{O}(\epsilon^{d-5})\,.
   \end{split}
\end{equation}
{With this additional term, the subleading divergent term for the CA conjecture in Ref.~\cite{Carmi:2016wjl} should be modified as}
\begin{equation}\label{subleadingCA}
  -\frac{\el^{d-1}}{16\pi^2\hbar}\int_B\td^{d-1}x\sqrt{\tilde{\sigma}^{(0)}}\left[\frac{4\tilde{K}^{(0)2}+4\tilde{K}_{ij}^{(0)}\tilde{K}^{(0)ij}-(3d+1)\tilde{R}^{(0)}+2(d+1)\tilde{\hat{R}}^{(0)}}{\epsilon^{d-3}(d-1)(d-2)(d-3)}\right]
\end{equation}
{so the first two divergent terms in Ref.~\cite{Reynolds:2016rvl} should be modified as}
\begin{equation}\label{subleadingCA2}
\begin{split}
\mathcal{C}_{\text{A,div}}&=\frac{\el^{d-1}}{4\pi^2\hbar\epsilon^{d-1}}\int_B\td^{d-1}x\sqrt{\tilde{\sigma}^{(0)}}\left[\ln(d-1)\left(1-\frac{\epsilon^2(\tilde{\hat{R}}^{(0)}-\tilde{R}^{(0)}/2)}{2(d-2)}\right)\right.\\
  &-\left.\epsilon^2\frac{d\tilde{K}^{(0)2}+2(d-1)\tilde{K}_{ij}^{(0)}\tilde{K}^{(0)ij}-3(d-1)\tilde{R}^{(0)}+2(d-1)\tilde{\hat{R}}^{(0)}}{2(d-1)(d-2)(d-3)}\right]+\mathcal{O}(\epsilon^{5-d})\,.
  \end{split}
\end{equation}

The result \eqref{subleadingCA2} can been obtained also by a different approach, in which we use the affinely parameterized $k^I$  i.e. $\kappa=0$. We still can write $k^I$ in the form shown in Eq.~\eqref{formkmu}, but we cannot demand that $\alpha$ is a constant. Instead, we assume $\alpha$ has the following series expansion with respective to $z$
\begin{equation}\label{expandalpha1}
  \alpha=\alpha^{(0)}+\alpha^{(1)}z^2+\cdots\,.
\end{equation}
Here except for $\alpha^{(0)}$, the other coefficients are only functions of $x^\mu$. By this approach, the null surface term is still zero but, {unlike the results in Refs.~\cite{Carmi:2016wjl,Reynolds:2016rvl}}, there is an additional contribution from $\alpha^{(1)}$. To see this, let us assume $\bar{k}_I$ is the affinely parameterized null normal vector for the other null surface at the joint. Then according to Ref.~\cite{Carmi:2016wjl}
\begin{equation}\label{formk2mu}
  \bar{k}_I=\beta(-dz-n_\mu\td x^\mu)\, ,
\end{equation}
where $\beta$ has a similar series expansion to $\alpha$:
\begin{equation}\label{expandbeta1}
  \beta=\beta^{(0)}+\beta^{(1)}z^2+\cdots \,.
\end{equation}
Thus the inner product of these two null vectors is\footnote{In Ref.~\cite{Carmi:2016wjl}, there is an order $\mathcal{O}(z^6)$ correction in Eq.~\eqref{k1k2a}. However, such correction is not necessary, as Eq.~\eqref{k1k2a} is an exact result in the FG coordinate system. }
\begin{equation}\label{k1k2a}
  k^I\bar{k}_I=2\frac{\alpha\beta}{\el^2} z^2\,.
\end{equation}
Using the expression in Eq.~\eqref{expressa}, we  find that
\begin{equation}\label{Ijointnew}
\begin{split}
  I_{\text{joint}}=&-\frac{\el^{d-1}}{8\pi\epsilon^{d-1}}\int_{\mathcal{J'}}\td^{d-1}x\sqrt{\tilde{\sigma}}\ln\left(\frac{\alpha\beta}{\el^2} \epsilon^2\right)\\
  =&-\frac{\el^{d-1}}{8\pi\epsilon^{d-1}}\int_{\mathcal{J'}}\td^{d-1}x\sqrt{\tilde{\sigma}}\ln\left\{\frac{\alpha^{(0)}\beta^{(0)}}{\el^2} \epsilon^2 \left[1+\left(\frac{\alpha^{(1)}}{\alpha^{(0)}}+\frac{\beta^{(1)}}{\beta^{(0)}}\right)\epsilon^2+\mathcal{O}(\epsilon^4)\right]\right\}\\
  =&-\frac{\el^{d-1}}{4\pi\epsilon^{d-1}}\ln\left(\frac{\sqrt{\alpha^{(0)}\beta^{(0)}}}{\el} \epsilon\right) \int_{\mathcal{J'}}\sqrt{\tilde{\sigma}}\td^{d-1}x \\ &-\frac{\el^{d-1}}{8\pi\epsilon^{d-3}}\int_{\mathcal{J'}}\td^{d-1}x\sqrt{\tilde{\sigma}^{(0)}}\left(\frac{\alpha^{(1)}}{\alpha^{(0)}}+\frac{\beta^{(1)}}{\beta^{(0)}}\right)+\mathcal{O}(\epsilon^{5-d})\,.
  \end{split}
\end{equation}
The logarithmic term in the last line of Eq.~\eqref{Ijointnew} is the same one in  Ref.~\cite{Carmi:2016wjl}, but there is an additional subleading divergent term due to $\alpha^{(1)}$ and $\beta^{(1)}$.

Now let us compute $\alpha^{(1)}$ and  $\beta^{(1)}$. Using the geodesic equation up to the order of $z^5$, we find that
\begin{equation}\label{solab1}
  \frac{\alpha^{(1)}}{\alpha^{(0)}}=\frac{\beta^{(1)}}{\beta^{(0)}}=\frac12\tilde{g}^{(1)}_{tt}\,.
\end{equation}
By this result, one can check that the subleading term should be
\begin{equation}\label{subleadingCA3}
  -\frac{\el^{d-1}}{8\pi^2\hbar}\int_\mathcal{B}\td^{d-1}x\sqrt{\tilde{\sigma}^{(0)}}\left[\frac{2\tilde{K}^{(0)2}+2\tilde{K}_{ij}^{(0)}\tilde{K}^{(0)ij}+(d^2-4d+1)\tilde{R}^{(0)}-d(d-3)\tilde{\hat{R}}^{(0)}}{\epsilon^{d-3}(d-1)(d-2)(d-3)}\right]\,,
\end{equation}
which is different from the result shown in Eq.~\eqref{subleadingCA}. It is because the action \eqref{actionull} without $I_\lambda$ depends on the parameterization of the null normal vector. Note that $\alpha^{(1)}$ and  $\beta^{(1)}$ also have additional contributions to the subleading term in $I_\lambda$. Using the method in Ref.~\cite{Reynolds:2016rvl}, one can compute such additional contribution. If we take both of two additional subleading contributions coming from $I_{\text{joint}}$ and $I_\lambda$ into account, we find the subleading divergent term is still the same as Eq.~\eqref{subleadingCA2}.

\section{The counterterms in higher dimension: examples in symmetric spaces}\label{app2}

Although the universal counterterms in higher dimension are complicated  in general, it is possible to obtain  simple formulas in some case: if the space has spherical, hyperbolic, or planar symmetry and the time slices at the boundary are given at constant $t$ ($t$ is the orbit of the timelike Killing vector field at the boundary).

Let us consider the metric of the form
\begin{equation}\label{metricfchi}
  \td s^2=-r^2f(r)e^{-\chi(r)}\td t^2+\frac{\td r^2}{r^2f(r)}+r^2\td\Sigma_{d-1,k}^2\,,
\end{equation}
where $k=1,0,-1$ for spherical, planar and hyperbolic space respectively.
The Ricci tensor  at any cut-off surface $r=r_m$ reads
\begin{equation}\label{Ricci-d}
  {R^\mu}_\nu=\text{diag}[0,(d-2)k/r_m^2,\cdots,(d-2)k/r_m^2]\,,
\end{equation}
The projection of the Ricci tensor on the constant time slices $t_L$ or $t_R$ is ${\hat{R}^i}_j=\text{diag}[(d-2)k/r_m^2,\cdots,(d-2)k/r_m^2]$. The extrinsic curvature of these two time slices at the cut-off surface vanish, i.e., $K_{ij}=0$. Thus, the scalar invariants $\sum_ic_{i,n}(d)[\mathcal{R},K]^{2n}_i$ are the combination of the $n$-th order scalar polynomials consisting of the contraction of ${R^\mu}_\nu$ or ${\hat{R}^i}_j$. Furthermore, in our case with the metric \eqref{metricfchi}, it is enough to consider the Ricci scalar $R$, because any other scalar invariants are equivalent to $R$.
Aa a result, $F_V^{(2n)}$ or $F_A^{(2n)}$ can be expressed as,
%
\begin{equation}\label{coeffcin1}
  \sum_ic_{i,n}(d)[\mathcal{R},K]^{2n}_i
  = c_{1,n} (d) R^{n}|_{r=r_m}\,,
\end{equation}
where we introduce $c_{1,n}$ without summation because we have only one kind of term in $[\mathcal{R},K]^{2n}_i$, the Ricci scalar $R$.
There is one exception which cannot be expressed as Eq. \eqref{coeffcin1}: if $d$ is odd and $n=(d-1)/2$, there is a logarithmic divergence, which should be treated separately following Eqs. \eqref{CAlog1} and \eqref{addlogCV}.

Our goal in the following subsections is to find the concrete expression of $c_{1,n}$ in both CV and CA conjectures so to find $F^{(2n)}_V$ and $F^{(2n)}_A$. There are two factors simplifying our analysis: i) the scalar curvature $R$ is coordinate independent so $c_{1,n}(d)$ is also coordinate independent. ii) with an
assumption that the matter part does not contribute the counterterms, $c_{1,n}(d)$ can be obtained from the vacuum solution.
%
%

\subsubsection*{CV conjecture}
The maximal volume with the boundary time slices $t_L=t_R=0$ is given by
\begin{equation}\label{CVvolum1}
  V=2\Sigma_{d-1,k}\int_{r_h}^{r_m}\frac{r^{d-2}}{\sqrt{f(r)}}\td r \,.
\end{equation}
The divergent structure can be obtained by setting $f(r)=1/\el^2+k/r^2$ and analysing the asymptotic behavior of
\begin{equation}\label{CVvolum2}
  2\Sigma_{d-1,k}\el\int_{r_h}^{r_m}r^{d-2}(1+k\el^2/r^2)^{-1/2}\td r \,.
\end{equation}
The divergent part is
\begin{equation*}\label{CVvolum3}
\begin{split}
  V_{\text{div}}&=2\Sigma_{d-1,k}\el\sum_{n=0}^{[\frac{d-1}2]}\int^{r_m}p_nk^{n}\el^{2n}r^{d-2-2n}\td r \\
  &=\left\{
  \begin{split}
  &2\Sigma_{d-1,k}\el\sum_{n=0}^{[\frac{d-1}2]}\frac{p_nk^{n}\el^{2n}}{d-1-2n}r_m^{d-1-2n}, \qquad \qquad \qquad \qquad \qquad \qquad  \qquad \text{(for\ even}~d)\,,\\
  &2\Sigma_{d-1,k}\el\sum_{n=0}^{[\frac{d-1}2]-1}\frac{p_nk^{n}\el^{2n}}{d-1-2n}r_m^{d-1-2n}+p_nk^{(d-1)/2}\el^{d-1}\ln(r_m/\el),~~~(\text{for odd}~d)\,,
  \end{split}
  \right.
  \end{split}
\end{equation*}
where the coefficients $p_n$ are defined by the following expansion
\begin{equation}\label{taylor}
  (1+x)^{-1/2}=\sum_{n=0}^{\infty}p_nx^n,~~~\text{with}~|x|<1\,,  \qquad p_n= \frac{\Gamma(1/2)}{\Gamma(n+1)\Gamma(1/2-n)}.
\end{equation}
Notice also that
\begin{equation}
 R|_{r=r_m} = \frac{(d-1)(d-2) k }{r_m^2} \,.
\end{equation}
Using the expression~\eqref{Ricci-d}, we can write the counterterms shown in Eq.~\eqref{expCVF}  as follows:
\begin{equation}\label{expCVF2}
\begin{split}
  V_{\text{ct}}&=\el \int_B\td^{d-1}x \ \sqrt{\sigma}\,  \sum_{n=0}^{[\frac{d-1}2]}  \el^{2n} c_n (d)R^{n} \,, 
  \end{split}
\end{equation}
where
\begin{equation}\label{valuesan}
  c_n(d)=\frac{p_n}{(d-1-2n)[(d-1)(d-2)]^n}  \,, 
\end{equation}
where  $p_n$ is given in Eq.~\eqref{taylor}.
%
In other words, $F_V^{(2n)}$ is written as
\begin{equation}\label{FVSAdS1}
  F_V^{(2n)}=\frac{\Gamma(\frac12)}{\Gamma(n+1)\Gamma(\frac12-n)}\frac{R^{n}}{(d-1-2n)[(d-1)(d-2)]^n} \,.
\end{equation}
When $d$ is odd and $n=(d-1)/2$, the counterterm is modified as \eqref{addlogCV}, where $\bar{F}_V^{(d-1)}=\frac{\Gamma(\frac12)}{\Gamma(n+1)\Gamma(\frac12-n)}\frac{R^{n}}{[(d-1)(d-2)]^n}$.

\subsubsection*{CA conjecture}

It is enough to find out the divergent structure of the on-shell action for the vacuum solution. The general results for the joint term and the boundary term can be obtained from the Eqs.~\eqref{SAdSjoint} and \eqref{IlambdaAdS}  with $\omega_{d-2}=0$:
\begin{equation}\label{SAdSjointbd}
\begin{split}
  I_{\text{joint}}+I_{\lambda}&=\frac{\Sigma_{d-1,k}}{2\pi}r_m^{d-1}\left[\ln(d-1)+\frac{1}{d-1}\right]+\frac{\Sigma_{d-1,k}}{4\pi}r_m^{d-1}\ln\left(1+\frac{k\el^2}{r_m^2}\right)\,.
  \end{split}
\end{equation}
The bulk term can be written as
\begin{equation}\label{SAdSCA0}
\begin{split}
  I_{\text{bulk}}&=\frac1{16\pi}\int_{\text{WDW}}\sqrt{-g}\td^{d+1}x\left[\mathfrak{R}+d(d-1)/\el^2\right]=-\frac{\Sigma_{d-1,k}d}{8\pi\el^2}\iint_{\text{WDW}}r^{d-1}\td t\td r\\
  &=-\frac{\Sigma_{d-1,k}}{8\pi\el^2}\iint_{\text{WDW}}(r^d)'\td t\td r=-\frac{\Sigma_{d-1,k}}{8\pi\el^2}\oint_{\partial\text{WDW}}r^d\td t\,.
  \end{split}
\end{equation}
where ``WDW'' means the ``Wheeler-DeWitt'' patch.
The divergent part in Eq.~\eqref{SAdSCA0} comes from the near boundary of the WDW patch, $\partial\text{WDW}$, which are the infalling and outgoing null geodesics coming from $t_L=t_R=0$ and $r=r_m$.  At these null geodesics, $t$ and $r$ satisfy $\td t\pm\td r/[r^2f(r)]=0$. Then we can see that the divergent part of $I_{\text{bulk}}$ can be expressed as,
\begin{equation}\label{SAdSCA}
\begin{split}
  I_{\text{bulk,div}}&=-\frac{\Sigma_{d-1,k}}{2\pi}\int^{r_m}r^{d-2}(1+k\el^2/r^2)^{-1}\td r\,.
  \end{split}
\end{equation}
After combining the Eqs.~\eqref{SAdSjointbd} and \eqref{SAdSCA}, we find that,
\begin{equation}\label{CAtotaldiv}
\begin{split}
  I_{\text{total,div}}&=\frac{\Sigma_{d-1,k}}{2\pi}r_m^{d-1}\ln(d-1)-\frac{\Sigma_{d-1,k}}{4\pi}\sum_{n=1}^{[\frac{d-1}2]}(-1)^nk^n\el^{2n}\frac{d-1}{n(d-1-2n)}r_m^{d-1-2n}\,.
  \end{split}
\end{equation}
When $d$ is odd, the last term in the summation in Eq.~\eqref{CAtotaldiv} ($2n=d-1$) should be replaced by a logarithmic term $(-1)^{(d-1)/2}k^{(d-1)/2}\el^{d-1}(d-1)\ln(r_m/\el)$. Comparing Eq. \eqref{CAtotaldiv} with Eq.~\eqref{CAdivn} and noting the all the divergent terms in Eq.~\eqref{CAtotaldiv} should be canceled by $2I_{\text{ct}}$, we find
%
%
that $F_A^{(0)}=\ln(d-1)/(4\pi)$ and,
\begin{equation}\label{FASAdS1}
  F_A^{(2n)}=-\frac1{8\pi}\frac{(-1)^n(d-1)R^{n}}{n(d-1-2n)[(d-1)(d-2)]^n}\,,\qquad \text{for}~n>0\,.
\end{equation}
When $d$ is odd and $n=(d-1)/2$, the counterterm is modified as \eqref{CAlog1}, where $\bar{F}_A^{(d-1)}=-\frac1{8\pi}\frac{(-1)^n(d-1)R^{n}}{n[(d-1)(d-2)]^n}$.

Eqs.~\eqref{FVSAdS1} and \eqref{FASAdS1} are the counterterms for any dimensions $d>2$. We used the specific coordinate but the final results are coordinate invariant.
 As a consistency check, we confirmed that we can reproduce the Eqs.~\eqref{surfSAdSA}, \eqref{subF2SAdS}, \eqref{CVplanar}, \eqref{CVknonzero} and \eqref{CVknonzero2} from  Eqs.~\eqref{FVSAdS1} and \eqref{FASAdS1}.

\bibliographystyle{JHEP}

\providecommand{\href}[2]{#2}\begingroup\raggedright\endgroup

\end{document}